\documentclass[aps,twocolumn,amsmath,amssymb,groupedaddress,prc]{revtex4}
\usepackage{graphicx}
\usepackage{simplewick}

\let\xxxhat\hat
\renewcommand{\hat}[1]{\boldsymbol {\xxxhat {#1}}}
\renewcommand{\vec}[1]{\boldsymbol{#1} }

\begin{document} 
\title{
Equation of state of low--density neutron matter and the
$\boldsymbol{ ^1S_0}$
pairing gap.
} 

\author{S.~Gandolfi} 
\affiliation{SISSA, International School of Advanced Studies,
via Beirut 2/4, 34014 Trieste, Italy}
\affiliation{INFN, Sezione di Trieste, Trieste, Italy}
\author{A. Yu.~Illarionov} 
\altaffiliation[Current address: ]{Dipartimento di Fisica, University of Trento,
via Sommarive 14, I-38050 Povo, Trento, Italy.}
\affiliation{SISSA, International School of Advanced Studies,
via Beirut 2/4, 34014 Trieste, Italy}
\affiliation{INFN, Sezione di Trieste, Trieste, Italy}
\author{F.~Pederiva}
\affiliation{Dipartimento di Fisica, University of Trento,
via Sommarive 14, I-38050 Povo, Trento, Italy}
\affiliation{INFN, Gruppo collegato di Trento, Trento, Italy}
\author{K. E.~Schmidt}
\affiliation{Department of Physics, Arizona State University,
Tempe, AZ, 85287, USA}
\author{S.~Fantoni}
\affiliation{SISSA, International School of Advanced Studies,
via Beirut 2/4, 34014 Trieste, Italy}
\affiliation{INFM/CNR--DEMOCRITOS National Simulation Center, Trieste,
Italy}
\affiliation{INFN, Sezione di Trieste, Trieste, Italy}

\begin{abstract}

We report results of the equation of state of neutron matter in the
low--density regime, where the Fermi wave vector ranges from
$0.4 ~{\rm fm}^{-1}  \leq k_F \leq 1.0 ~{\rm fm}^{-1}$.
Neutron matter in this
regime is superfluid because of the strong and attractive interaction
in the $^1S_0$ channel. The properties of this superfluid matter
are calculated 
starting from a realistic Hamiltonian that contains modern two--
and three--body interactions. The ground state energy and the
$^1S_0$ superfluid energy gap are calculated using
the Auxiliary Field Diffusion Monte Carlo method.
We study the structure
of the ground state by looking at
pair distribution functions as well as the
Cooper-pair wave function used in the calculations.
\end{abstract}

\maketitle

\section{Introduction}
\label{sec:intro}
Pure neutron matter is the natural first approximation to the baryonic matter
that composes the bulk of neutron stars.
At very low densities below neutron drip, (i.e. where the
Fermi wave vector is roughly, $k_F \lesssim 0.2$ ${\rm fm}^{-1}$)
neutron star matter is conjectured
to be nuclei surrounded by a relativistic gas of electrons\cite{pethick95a}.
At higher densities the matter becomes liquid and very neutron rich.
Here we study matter at Fermi wave vector
$0.4 ~{\rm fm}^{-1} \leq k_F \leq 1.0 ~{\rm fm}^{-1}$
where it is reasonable to approximate it as pure neutron matter, and also
extend some of our results into the lower density regime in order to compare
with other calculations.

At these densities
the interaction is dominated by the $^1S_0$ channel
with a large and negative scattering length, $a \simeq -18.5 ~{\rm fm}$.
The product of the effective range and the Fermi wave vector is
of order unity, so, while the form of interaction cannot be neglected,
it becomes less important.
Analysis of the phase shifts of the neutron-neutron $^1S_0$ interaction
indicates that neutrons should pair and form a superfluid. Therefore the
superfluid phase must be included when investigating the equation of
state in this regime.

Many methods have been used to approximately calculate the equation of state.
One class of methods uses
Skyrme or relativistic mean-field methods that use effective interactions
that have been fit to the properties of nuclei. However, even those
calculations that
describe neutron-rich nuclei reasonably well, give rather different
equations of state for pure neutron matter\cite{heiselberg00}.
We instead use a nonrelativistic Hamiltonian with two- and three-body
interactions. All modern accurate two-body interactions fit the Nijmegen
data\cite{stoks93} within experimental errors and should give
essentially the same equation of state at low density. At longer range,
these interactions are dominated
by the one-pion exchange and have strong spin-isospin dependence, which
must be included for accurate predictions.
Three- and higher-body interactions
are less well known but in this density regime they are small.

Our calculations
extend the work we first reported in ref.~\cite{gandolfi08b}.
The ground state of neutron matter is computed using
the auxiliary field diffusion
Monte Carlo \cite{schmidt99} (AFDMC) algorithm; it is an extension of
the diffusion Monte Carlo method\cite{anderson75}, and
Green's function Monte Carlo method\cite{carlson87}.
These Monte Carlo algorithms are very well
suited to project a trial wave function onto the ground state in order
to study the ground state properties of a system.
The Green's
function Monte Carlo method has been used
to study the properties of light nuclei with very
high accuracy\cite{pieper05}.
The advantage of the auxiliary field diffusion Monte Carlo method over
the Green's function Monte Carlo method is that it
can be extended
to larger nuclear systems; in fact it has been used
to calculate properties of heavy nuclei\cite{gandolfi07b}, neutron-rich
isotopes\cite{gandolfi06,gandolfi08} and neutron\cite{sarsa03,gandolfi09}
and nuclear matter\cite{gandolfi07} by simulating systems with upwards of
one hundred nucleons.

The equation of state of neutron matter in the low-density regime has
been a subject of many previous calculations\cite{friedman81,akmal98,
carlson03,schwenk05,borasoy08,lee09,epelbaum09,gezerlis08,gandolfi08b,
abe09,abe09b}.
While
in this regime, different Hamiltonians and different methods give
similar behavior for the energy as a function of the density, there are
appreciable differences in other important properties.
In particular,
the value of the $^1S_0$ superfluid energy gap is at present not well clarified
and strongly depends on the Hamiltonian and the solution
method\cite{gandolfi08b}.
In this paper we focus on both the energy and the energy gap by
considering a fully realistic Hamiltonian, and solve for the ground state
using the AFDMC technique.  As a starting point for the calculation
we considered two forms for the trial wave function. The first is
a filled Fermi sea  having the properties of a normal Fermi liquid which
we will call the normal phase. The
second has neutrons paired in the $^1S_0$ channel with a
Bardeen-Cooper-Schrieffer (BCS)
superfluid structure\cite{bardeen57}.
At a fixed density, we find that the
superfluid
phase of the system is only marginally favored compared
to the normal phase. However, to calculate the superfluid energy gap,
the BCS structure must be used.

\section{Hamiltonian}
\label{sec:hamiltonian}
We study
the ground state of neutron matter beginning with
the non--relativistic nuclear Hamiltonian
\begin{equation}
\label{hamiltonian}
H=-\frac{\hbar^2}{2m}\sum_{i=1}^N\nabla_i^2+\sum_{i<j}v_{ij}
+\sum_{i<j<k}V_{ijk} \,,
\end{equation}
where $m$ is the mass of the neutron, and $v_{ij}$ and $V_{ijk}$ are two--
and three--body potentials.  Such a form for the Hamiltonian (with the
kinetic energy modified to take into account the mass difference of
the neutron and proton) has been
shown to describe properties of light nuclei in a good agreement with
experimental data (see ref.~\cite{pieper05} and references therein).
All the degrees of freedom responsible for the interaction between
nucleons (such as the $\pi$, $\rho$, $\Delta$, etc.) are integrated out
and included in $v_{ij}$ and $V_{ijk}$.

At present, several realistic two-nucleon interactions fit
scattering data with very high precision. We use the two-nucleon
potentials
belonging to the Argonne family~\cite{wiringa95}.  Such interactions
are written as
\begin{equation}
\label{V2body}
v_{ij} = \sum_{p=1}^{M} v_p(r_{ij}) O^{(p)}(i,j)\,,
\end{equation}
where $O^{(p)}(i,j)$ are spin--isospin dependent operators.  The number
of operators $M$ characterizes the interaction; the most accurate for
the Argonne family
is the Argonne AV18 with M=18~\cite{wiringa95}. Here we consider a
simpler form derived from AV18, namely the AV8$^\prime$~\cite{wiringa02}
with a smaller number of operators.  For many systems, the difference
between this simpler form and the full AV18 potential can be computed
perturbatively~\cite{pudliner97,gandolfi09}, as has been done in all Green's function
Monte Carlo calculations to date. Most of the contribution of the two-nucleon
interaction is
due to one-pion exchange between nucleons, but the effect of other
mesons exchanges as well as some phenomenological terms are also included.

The eight $O^{(p)}(i,j)$ operators in AV8$^\prime$ are given by the four central
components $1$, $\vec\tau_i\cdot\vec\tau_j$, $\vec  \sigma_i \cdot \vec
\sigma_j$, $(\vec \sigma_i \cdot \vec \sigma_j)(\vec \tau_i\cdot \vec
\tau_j)$, the tensor $S_{ij}$, the tensor--$\tau$ component $S_{ij}
\vec \tau_i\cdot \vec \tau_j$, where $S_{ij} = 3 (\vec \sigma_i \cdot
\hat r_{ij}) (\vec \sigma_j \cdot \hat r_{ij}) -\vec \sigma_i \cdot
\vec \sigma_j$, the spin--orbit $\vec L_{ij}\cdot\vec S_{ij}$ and the
spin--orbit--$\tau$ $(\vec L_{ij}\cdot\vec S_{ij})(\vec\tau_i\cdot
\vec \tau_j)$, where $\vec L_{ij}$ and $\vec S_{ij}$ are the relative
angular momentum and the total spin of the pair $ij$.  All the parameters
describing the radial functions of each operator in AV18 are fit to
nucleon-nucleon scattering data below 350 MeV in the
Nijmegen database~\cite{stoks93}.
The AV8$^\prime$ interaction is obtained by starting from AV18 and making
an isoscalar projection.
It is refit in order to keep the most important features of AV18
in the scattering data and the properties of the deuteron~\cite{wiringa02}.

The three--nucleon interaction is essential to overcome the
underbinding of nuclei with more than two nucleons. While the two-nucleon
interaction
is fit
to scattering data and correctly gives the deuteron binding energy,
it is not sufficient to describe the ground state
of light nuclei with three or more nucleons.
The Urbana-IX (UIX) potential corrects this,
and was fit to obtain the correct triton energy using Green's function
Monte Carlo and to correctly reproduce
the expected saturation energy of nuclear matter within the
Fermi hypernetted-chain approximation~\cite{pudliner95}.
It contains a Fujita--Miyazawa term~\cite{fujita57}
that describes the exchange of two pions between three nucleons,
with the creation of an intermediate excited $\Delta$ state.  Again, a
phenomenological part is added to sum all the other neglected terms.
The generic form of UIX is:
\begin{equation}
\label{V3body}
V_{ijk}=V_{2\pi}+V_R \,.
\end{equation}
The Fujita-Miyazawa term~\cite{fujita57} is spin--isospin dependent:
\begin{align}
\label{V2pi}
V_{2\pi}=A_{2\pi}\sum_{cyc}\Big[&\{X_{ij},X_{jk}\}
\{\vec \tau_i\cdot\vec \tau_j,\vec \tau_j\cdot\vec \tau_k\}+
\nonumber \\
\dfrac{1}{4}&[X_{ij},X_{jk}]
[\vec \tau_i\cdot\vec \tau_j,\vec \tau_j\cdot\vec \tau_k]\Big] \,,
\end{align}
where the $X_{ij}$ operators describe the one pion exchange (see
ref. \cite{pieper01} for details).
The phenomenological part of UIX is
\begin{equation}
\label{V3b_R}
V_{ijk}^R=U_0 \sum_{cyc}T^2(m_\pi r_{ij})T^2(m_\pi r_{jk}) \,.
\end{equation}
The factors $A_{2\pi}$ and $U_0$ are kept as fitting parameters.
Other forms of three-nucleon interaction,
called the Illinois
forces~\cite{pieper01}, which includes three--nucleon Feynman diagrams
with two-$\Delta$ intermediate states, are available.  Unfortunately
they provide unrealistic overbinding of neutron systems when the density
increases~\cite{gandolfi07c,sarsa03} and they do not seem to describe,
realistically, higher density (i.e. $\rho\ge\rho_0=0.16$ fm$^{-3}$) nucleonic
systems.  However in the low--density regime considered in this paper,
the contribution of the three--body interaction is very small
compared to the total energy of the system, so that the small
errors in the UIX interaction should have negligible contributions to the
equation of state and energy gap.

\section{AFDMC method and the pfaffian wave function}
\label{sec:method}

Uniform neutron matter is simulated by solving the ground state of a
fixed number $N$ of neutrons in a periodic box, whose volume is fixed by
the density of the system.  The ground state of the system is calculated
by means of the AFDMC algorithm~\cite{schmidt99}.
Diffusion Monte Carlo projects
out the lowest-energy state from a trial wave function $\psi_T$ by a
propagation in imaginary time:
\begin{equation}
\psi(\tau)=e^{-(H-E_T)\tau}\psi_T \,,
\end{equation}
where $E_T$ is a normalization factor. In the
$\tau\rightarrow\infty$ limit the only component 
of $\psi_T$ that survives is the lowest-energy one not orthogonal to $\psi_T$:
\begin{equation}
\phi_0=\lim_{\tau\rightarrow\infty}\psi(\tau) \,.
\end{equation}
The evolution in imaginary time is performed by solving the integral equation
\begin{equation}
\psi(R,\tau)=\int dR^\prime G(R,R^\prime,\tau) \psi_T(R^\prime) \,,
\end{equation}
where $G(R,R^\prime,\tau)$ is the Green's function of the Hamiltonian that
contains a diffusion term, coming from the kinetic operator in $H$, and a
branching term from the potential. The exact form of $G(R,R^\prime,\tau)$
is unknown, but it can be accurately approximated in the limit of
$\Delta\tau \to 0$. The above integral equation is then solved
iteratively, with a small time step,
for a sufficiently large number of steps. A detailed description
of the algorithm as well as the importance sampling technique used to
reduce the variance can be found in \cite{guardiola98,mitas99}.

The presence of spin operators in the Hamiltonian requires a summation of
all possible good spin states in the wave function~\cite{carlson99b}. This
summation grows exponentially with the number of neutrons;
for example, for a system of 14 neutrons the computation of
$\langle\psi(R)\vert\psi(R)\rangle$ is a sum of squares of $2^{14}$
spin amplitudes. The explicit summation of spin states is performed
in Green's function Monte Carlo, but not in 
AFDMC, calculations where the spin states are
sampled using Monte Carlo techniques~\cite{schmidt99}.  This sampling
is performed by
reducing the quadratic dependence of spin operators in the exponential
to a linear form
by means of the Hubbard-Stratonovich transformation.
The effect of an exponential of a linear combination of spin operators consists
of a rotation of the spinor for each neutron during the propagation.
In order to have an efficient algorithm, the trial function must be
chosen so that it can be efficiently evaluated when each neutron
is in a specific position and spinor state.

Since both positions and spins can be sampled,
the AFDMC method can be used to solve for the ground state of much larger
systems -- more than one-hundred neutrons --
than Green's function Monte Carlo with full spin summations.

More detailed explanations of the AFDMC method and how to include the
full two- and three-nucleon interactions
in the propagator can be found in
refs.~\cite{sarsa03,pederiva04,gandolfi07c,gandolfi09}, where
the fixed-phase approximation used to control the fermion
sign problem is also discussed.

The AFDMC method projects out the lowest energy state with the same symmetry
as the trial wave function from which the projection is started.
The general form of the trial wave function is
\begin{equation}
\label{psi_T}
\psi_T
(R,S)
=\left[\prod_{i<j}f_J(r_{ij})\right] \Phi(R,S) \,,
\end{equation}
where $R\equiv (\vec r_1,\dots,\vec r_N) $ represents the spatial
coordinates and $S\equiv (s_1,\dots ,s_N)$ the spin states of the
neutrons.  The spin assignments $s_i$ consist of giving the two spinor
components for each neutron, namely the two complex numbers $a_i$, $b_i$
where
\begin{equation}
\label{eq:spinor}
\vert s_i\rangle=a_i\lvert\uparrow\rangle+b_i\lvert\downarrow\rangle \,,
\end{equation}
and the
$\{\lvert\uparrow\rangle,\lvert\downarrow\rangle\}$ is the spin-up and
spin-down basis. The function $f_J$ entering in the so called Jastrow part of the
trial wave function has only the role of reducing the
overlap of neutrons and thereby reducing the energy variance.  Since it
does not change the phase of the wave function, it does not influence
the computed energy value in projections methods. The function $f_J$
is computed as described in ref. \cite{gandolfi09}.

The antisymmetric part $\Phi$ of the trial wave function is usually given
by the ground state of non--interacting Fermions (Fermi gas), which is
written as a Slater determinant of single particle functions. For example
homogeneous systems are usually simulated by considering plane--waves as
orbitals. In this case
\begin{equation} 
\Phi_{n}(R,S) ={\cal A} \left[\phi_1(\vec r_1,s_1)\dots
\phi_N(\vec r_N,s_N)\right] \,,
\end{equation}
where ${\cal A}$ is the antisymmetrizer (see Eq. \ref{eq:antisym}),
\begin{equation}
\phi_\alpha(\vec r_i,s_i)=e^{i\vec k_\alpha\cdot\vec r_i}
\langle s_i \vert \chi_{s,m_s,\alpha} \rangle \,,
\end{equation}
and $\alpha$ is the set of quantum numbers of single-particle orbitals
that are plane waves fitting the box.
The correct symmetry of the ground state is given using 
the closed shells occurring when the total number of Fermions in a particular
spin configuration is 1, 7, 19, 27, 33,...

However, in superfluid neutron matter there is a strong coupling
between neutrons, and a wave function heaving a BCS structure must be
used.

BCS pairing correlations can substantially change the nodal structure of a
trial wave function~\cite{carlson03c,chang04b}.  This change, which gives
the off-diagonal long-range order of the superfluid phase, will greatly
alter the fixed-phase (or constrained path) energy. In order to correctly
describe the superfluid ground state with these quantum Monte Carlo methods,
we need to use a trial wave function with explicit pairing.  For central
potentials and singlet pairing, the BCS trial function can be written as a
determinant~\cite{bouchaud1988,carlson03c}.
However, for problems with a tensor
force, or for spin triplet pairing, a general pairing state must be used.

A fully paired state of $N$ neutrons
can be written, as shown in appendix \ref{appendixa},
as
\begin{equation}
\mathcal{A} \left[ \phi_{12} \phi_{34} \ldots \phi_{N-1,N} \right ]\,,
\end{equation}
Similarly, we can construct
a general state with $n$ paired and $o$ unpaired orbitals for a total of
$N = 2n+o$ particles as
\begin{equation}
\mathcal{A} \left[ \phi_{12} \phi_{34} \ldots \phi_{2n-1,2n} \ldots
                    \psi_1(2n+1) \ldots \psi_o(N) \right] \ ,
\label{2n+o}
\end{equation}
which is the pfaffian of the $(N+o) \times (N+o)$ skew-symmetric
matrix~\cite{bouchaud1988}
\begin{widetext}
\begin{equation}
 \left (
\begin{array}{cccccccc}
0 & \phi_{12} & \phi_{13} & \ldots & \phi_{1N}
  & \psi_1(1) & \ldots & \psi_o(1) \\
-\phi_{12} & 0 & \phi_{23} & \ldots & \phi_{2N}
  & \psi_1(2) & \ldots & \psi_o(2) \\
\vdots & \vdots & \ddots & \vdots & \vdots & \vdots & \vdots & \vdots \\
-\phi_{1N} & -\phi_{2N} & -\phi_{3N} & \ldots & 0
  & \psi_1(N) & \ldots & \psi_o(N) \\
-\psi_1(1) & -\psi_1(2) & -\psi_1(3) & \ldots & -\psi_1(N)
  & 0 & \ldots & 0 \\
\vdots & \vdots & \vdots & \vdots & \vdots & \vdots & \ddots & \vdots \\
-\psi_o(1) & -\psi_o(2) & -\psi_o(3) & \ldots & -\psi_o(N)
  & 0 & \ldots & 0 \\
\end{array}
\right ) \ ,
\label{N:pfaf}
\end{equation}
\end{widetext}
where the lower $o \times o$ section is all zeroes.

The pfaffian is the antisymmetric product
\begin{equation}
\label{eq1}
{\rm Pf} A = {\cal A} [ a_{12} a_{34} a_{56}... a_{N-1,N} ] \,.
\end{equation}
The result is
normalized such that every equivalent term occurs only once, and
$a_{ij} = -a_{ji}$.

Just as the determinant of a dense matrix can be calculated efficiently
in order $N^3$ operations, similar elimination methods can compute
the pfaffian. The basic pfaffian calculational
methods we use here, and have used, for all previous
superfluid neutron matter studies\cite{fabrocini05,gandolfi08b},
are described in some detail in section II of ~\cite{bajdich08},
and those results are summarized
in appendix \ref{appendixb} along with some additional techniques
needed for these nuclear calculations.

The nuclear Hamiltonian has spin-dependent terms that can flip the
spin. For the simpler case of a purely central potential, the Hamiltonian
will not change the particles' spin. Therefore in this simpler
case we can solve for the ground state in one sector where each particle
has a specified spin, and we only need to antisymmetrize over the particles
with the same spin. In that case, $\Phi_{BCS}$ reduces to a determinant.
Since in our AFDMC method, the Hamiltonian can change the particles' spin,
and the particles can then take on any spinor value,
we need to be able to evaluate the trial wave function
for arbitrary spinor values for each particle. Therefore the pfaffian
which gives the full antisymmetric form must be used. As shown in
appendix \ref{appendixa},
the pairing orbitals $\phi$ we used have the form
\begin{align}
\phi(\vec r_{ij},s_i,s_j)&= \sum_\alpha\frac{v_{k_\alpha}}{u_{k_\alpha}}
e^{i \vec k_\alpha\cdot\vec r_{ij}}\chi(s_i,s_j)
\nonumber \\
&= \sum_\alpha c_\alpha e^{i \vec k_\alpha\cdot\vec r_{ij}}\chi(s_i,s_j)
\,,
\label{eq:orbitals}
\end{align}
where the sum over $\alpha$ indicates the $k$-space shells of the
cube with $\vec k$ values
\begin{equation}
k_{n_xn_yn_z} = \frac{2\pi}{L}
 (n_x \hat x + n_y \hat y + n_z \hat z)
\end{equation}
for integer $n_x$, $n_y$, and $n_z$.
The function $\chi$ is the spin-singlet wave function for two neutrons
\begin{equation}
\chi(s_i,s_j)= \frac{1}{\sqrt{2}}
\left(\langle s_i s_j |\uparrow \downarrow\rangle
-\langle s_i s_j |\downarrow \uparrow\rangle\right) \,.
\end{equation}
With the spin states given as spinors as in Eq. \ref{eq:spinor}
this becomes
\begin{equation}
\chi(s_i,s_j) = \frac{a_i^* b_j^*-b_i^*a_j^*}{\sqrt{2}}
\end{equation}
Note that if the pairing coefficients
$c_\alpha$ are zero for all $|\vec k_\alpha| > k_F$, the pfaffian
of Eq.~\ref{N:pfaf} is exactly the Slater
determinant of spin up and down neutrons filling the Fermi sea, and
the pfaffian form goes over to the normal liquid state.
The parameters $c_\alpha$ are chosen variationally by performing a correlated
basis function
calculation~\cite{fabrocini05,fabrocini08}.  However, various other
wave functions were considered to ascertain the effect of a
particular choice on the results.

\section{Results}
\label{sec:results}

\subsection{Equation of State}
\label{subsec:EOS}

We computed the energy of neutron matter by simulating neutrons in a
periodic box at densities corresponding to $k_F = 0.4$, $0.6$, $0.8$
and $1.0$ fm$^{-1}$ using in the trial wave function both $\Phi_{n}$
and $\Phi_{BCS}$.
We found that the absolute energy is slightly different
depending on the
choice of the trial function $\Phi$.
The results obtained using the two different trial wave functions are reported
in table~\ref{tab:envsebcs}.
\begin{table}
\begin{center}
\begin{tabular}{cc|cc}
\hline
$k_F$ [fm$^{-1}$] & $\rho$ [fm$^{-3}$] & $E_{n}/N$ & $E_{BCS}/N$ \\
\hline
0.4               & 0.00216 &            1.289(2) & 1.239(2) \\
0.6               & 0.00730 &            2.606(4) & 2.579(2) \\
0.8               & 0.01729 &            4.277(7) & 4.305(3) \\
1.0               & 0.03377 &            6.197(2) & 6.231(3) \\
\hline
\end{tabular}
\end{center}
\caption{AFDMC energies per particle for 66 neutrons interacting with
the AV8$^\prime$+UIX interaction in a periodic box as a function of
the Fermi wave vector and corresponding density $\rho$. 
The values $E_{n}$ correspond to the simulation
of neutron matter using the Fermi gas ground state in the trial wave
function, while $E_{BCS}$ are the results obtained using $\Phi_{BCS}$.
All the energies are expressed in MeV.}
\label{tab:envsebcs}
\end{table}
As can be seen, the BCS state is favored
at $k_F=0.4$ and $0.6$~fm$^{-1}$, while the normal state trial function
gives the lowest energy at $k_F=0.8$ and $1.0$~fm$^{-1}$.
The maximum difference between the results for the
two different trial wave functions is
about 4 percent of the total energy at $k_F=0.4$~fm$^{-1}$, probably because
at such a low density the pairing between neutrons in the $^1S_0$ channel is very
important and $\Phi_{BCS}$ includes such correlations in the wave function
in a more effective way.  In the other cases the energies obtained with
$\Phi_{BCS}$ and $\Phi_{n}$ are within 1 percent.

Since the coefficients entering in $\Phi_{BCS}$ were chosen by
a correlated basis function
calculation that adds a two body correlation factor to the usual
BCS state~\cite{fabrocini05,fabrocini08}, in order to determine if this
method is adequate for finding a good BCS form,
we repeated some of the calculations
using different coefficients.  In particular we tried using, as a pairing
function, the solution from the uncorrelated BCS equation, as well
as a pairing
function with the same form as that of ref.~\cite{gezerlis08,gezerlis08b}.
This calculation has carefully optimized coefficients, but the
interaction is the $^1S_0$ channel of AV18 acting only between
unlike spins.
In both cases we find the energy is slightly higher than that found
when using the correlated basis function coefficients.

\begin{figure}[ht]
\vspace{1cm}
\begin{center}
\includegraphics[width=8cm]{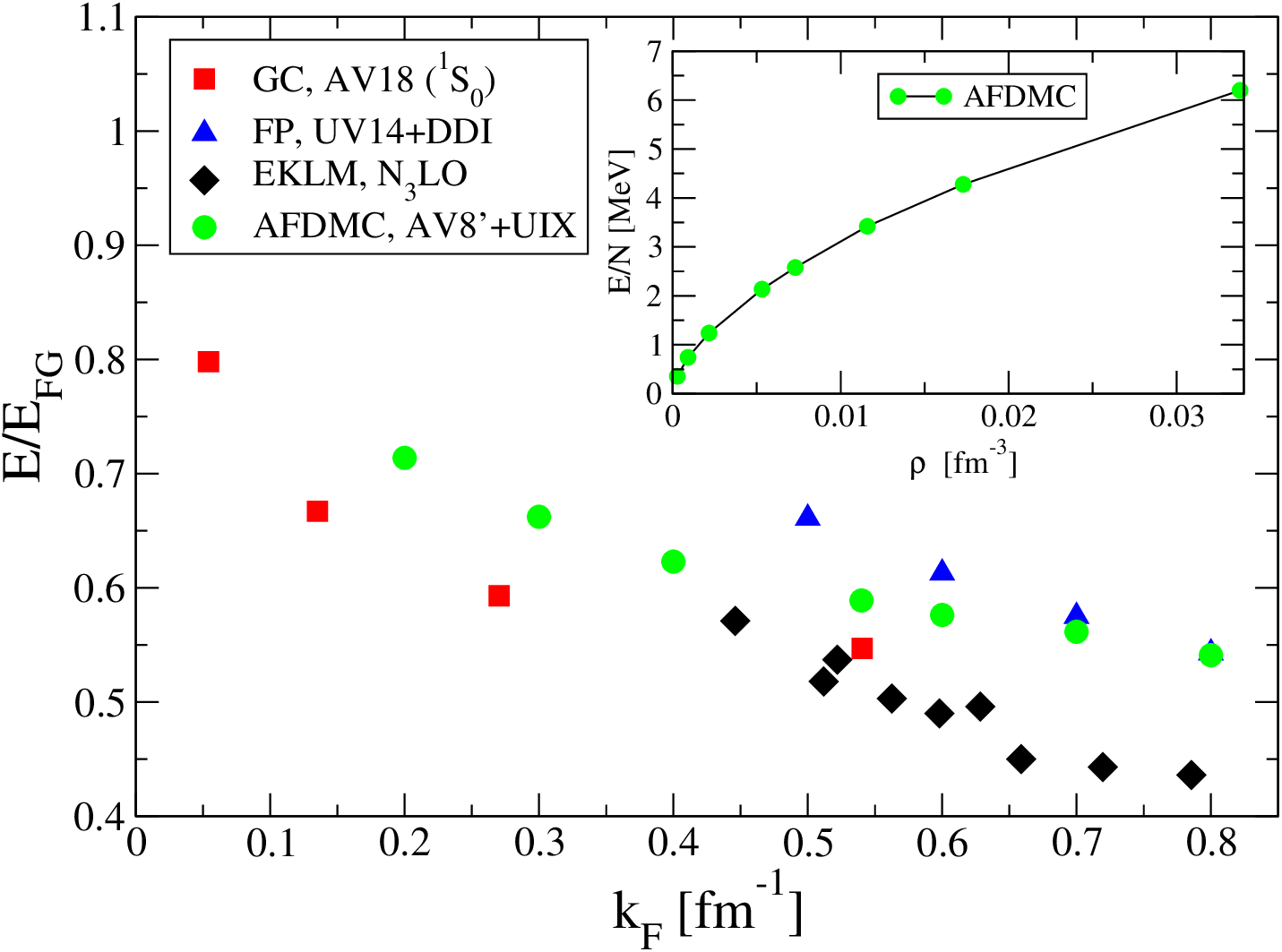}
\vspace{0.5cm}
\caption{(color online) The equation of state of neutron matter as a function
of the Fermi wave vector $k_F$. The energy has been divided by the energy
of the
noninteracting Fermi gas, $E_{FG}=\frac{3}{5}\frac{\hbar^2 k_F^2}{2m}N$. 
The AFDMC result is obtained using the full
Hamiltonian AV8$^\prime$+UIX (green circles), and compared with the
results of Gezerlis and Carlson (red squares) who considered a simpler
Hamiltonian~\cite{gezerlis08}. The blue triangles correspond to the
calculation of Friedman and Pandharipande using the Urbana $v_{14}$
two-nucleon interaction
modified to include some three--body effects~\cite{friedman81}. 
The black diamonds show the results of Epelbaum, Krebs, Lee and 
Mei{\ss}ner\cite{epelbaum09}.
In the
inner part of the figure, the AFDMC energy in MeV is shown
as a function of the
density $\rho$ in fm$^{-3}$ is also displayed along with a curve
to guide the eye.}
\label{fig:eos}
\end{center}
\end{figure}

The equation of state
of low--density neutron matter, computed using $\Phi_{BCS}$ is
displayed in Fig.~\ref{fig:eos}, compared to the diffusion
Monte Carlo results of Gezerlis
and Carlson~\cite{gezerlis08}, to the variational cluster summation
calculation of Friedman and Pandharipande~\cite{friedman81} and to
the results of Epelbaum, Krebs, Lee and Mei{\ss}ner\cite{epelbaum09}.
The differences between the various calculations are due to different
approximations and interactions used.
The AFDMC method uses a realistic Hamiltonian containing
a modern two--body and the corresponding three--body force. The variational
cluster summation calculation was performed
using the older Urbana $v_{14}$ two-nucleon interaction\cite{lagaris81}
modified to include
a density
dependent term that models the effect of a three--body force. As mentioned
above, the
calculation of Gezerlis and Carlson
uses only the
$^1S_0$ channel interaction of AV18 between unlike spins. This choice is
motivated by the fact that this channel is dominant in neutron matter in
this regime. However the effect of other channels 
as well as using the $^1S_0$ interaction partly in the
triplet channel, since all unlike spin pairs interact,
could play an important role in the many--body
correlations of the system.  Finally, Epelbaum and collaborators computed
the equation of state
within the chiral effective field theory by simulating neutrons
on the lattice up to the N$_3$LO order\cite{epelbaum09}.

Each of these calculations used different methods
to solve for
the ground state.  Both the AFDMC and the diffusion
Monte Carlo method used by Gezerlis and Carlson are
projection methods that, apart from the constraint used to control the Fermion
sign problem, are exact. However, the constraint plays an important
role in finding the correct ground state, and different trial functions
give different constraints and therefore different results. For these
two Monte Carlo methods, trial functions have the same kind of BCS form.
However Gezerlis and Carlson use a different approach for
the choice of the coefficients entering in the pairing orbitals of
Eq.~\ref{eq:orbitals}. Their
$c_\alpha$ parameters
are chosen by varying them to minimize the fixed--node
energy~\cite{carlson03c,chang04b,gezerlis08}. Unfortunately this same
technique
is not currently applicable to
AFDMC because the variance of the calculation is
too high to be able to choose the coefficients in a reasonable amount of
computational time. The $c_\alpha$ used
in our AFDMC calculations
are chosen instead by using a correlated BCS wave function solved within
the CBF/BCS theory~\cite{fabrocini05,fabrocini08} as discussed
above.  The variational
cluster summation calculation may suffer from important uncontrolled
approximations coming from the cluster expansion as we recently pointed
out in our paper comparing the equation of state
of neutron matter at higher
densities~\cite{gandolfi09}. In addition, the variational
cluster summation calculation does
not include any pairing correlations in the variational wave function.
The equation of state
can be computed using N$_3$LO as described
in~\cite{borasoy08,lee09,epelbaum09}, and the results are available
for a small number of neutrons (N=12). They predict an
equation of state
that is globally lower than the other results.
This model, while very promising because it attacks
the problem from a more fundamental point of view,
will need to be extended to larger systems.

\subsection{Superfluid gap}
\label{subsec:gap}

In a full many--body calculation the superfluid gap can be evaluated by 
using the difference:
\begin{equation}
\label{eq:gap}
\Delta(N) = E(N)-\dfrac{1}{2}\left[E(N+1) + E(N-1)\right] \,,
\end{equation}
where the number of neutrons $N$ is taken to be odd.  The AFDMC
algorithm can be used to simulate very large systems with up to a hundred
nucleons~\cite{sarsa03,gandolfi07,gandolfi07c,gandolfi09}. Unfortunately, because
the gap has to be evaluated as the difference between total energies
of different systems, the statistical error related to $\Delta$
is proportional to the number of neutrons, and we have not been able
to develop an efficient method of correlated sampling.  As a consequence, in
principle, the number of neutrons is arbitrary but if $N$ is too large,
the statistical error affecting the gap becomes larger than the
gap itself. The maximum number of neutrons used in this work is 68.

Particular care was taken to check that the AFDMC had converged.
The simulations were repeated with different time steps.
Neither the energy nor the gap is
dependent on the time step used -- the extrapolation to the zero
limit is within our error bars.

The gap is strongly dependent on the number of neutrons for
small N.
For both $k_F=0.4$ and $0.6$ fm$^{-1}$
the $\Delta$ computed with $N=12\dots18$ is noticeably larger compared
to that computed with $N=62\dots68$. We find that at $k_F=0.4$
fm$^{-1}$ the gap is $\Delta(14)=1.79(6)$ MeV and $\Delta(66)=1.5(2)$ MeV, while
at $k_F=0.6$ fm$^{-1}$ $\Delta(14)=2.59(6)$ MeV and $\Delta(66)=2.1(2)$ MeV.
This behavior is well described by the analysis of Gezerlis and Carlson
who
solved the BCS equation in the simulation cell, and then reproduced this
effect by using
diffusion Monte Carlo~\cite{gezerlis08}. In their paper they calculate
with up to 90 particles without observing a substantial
change in
the gap compared to that given by simulating the system with about
66 particles, giving us confidence that our gaps have converged.

\begin{figure}[ht]
\vspace{1cm}
\begin{center}
\includegraphics[width=8cm]{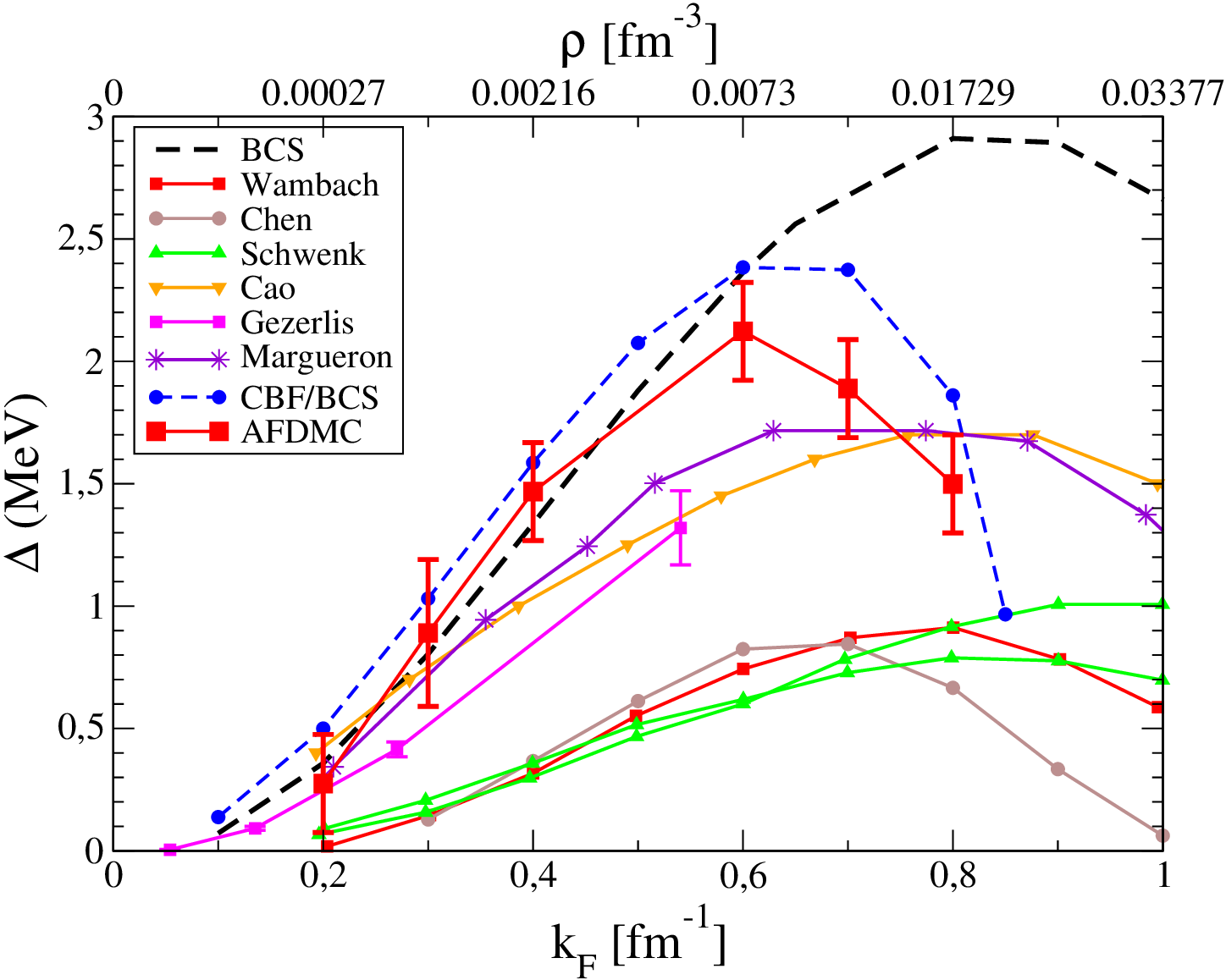}
\vspace{0.5cm}
\caption{(color online) The $^1S_0$ pairing gap of neutron matter as a
function of the Fermi wave vector $k_F$ computed with different methods. In
the figure we display works of Wambach et al.~\cite{wambach93},
Chen et al.~\cite{chen93}, Schulze et al.~\cite{schulze96}, Schwenk
et al.~\cite{schwenk03}, Cao et al.~\cite{cao06}, Gezerlis and
Carlson~\cite{gezerlis08} and Margueron et al.~\cite{margueron08}. All
the results are compared with a BCS calculation (dashed line).}
\label{fig:gap}
\end{center}
\end{figure}

We report in Fig.~\ref{fig:gap} the superfluid gap computed
with AFDMC using $N=62\dots68$, compared with other
calculations. It is clear that the different methods used to compute
the pairing gap give different results.  
The mean--field BCS result is essentially unchanged when other realistic 
two-nucleon
interactions are used~\cite{fabrocini08,hebeler07}, and reaches a maximum of about
3 MeV. This is because all the two--body interaction are
fit by reproducing the S-- and P--wave components from experimental
data. However a realistic study of the pairing gap must include
the corrections due to the polarization effects given by the medium.
The various results can be essentially divided in two different groups
according to the different way used to include this effect: the
many--body calculation using effective interactions based on Brueckner
theory or Hartree--Fock calculations, or the microscopic calculations
(Monte
Carlo methods or correlated basis function theory)
where the whole Hamiltonian describing the system is solved.
The many--body effective-interaction
calculations of Wambach et al.~\cite{wambach93}, Chen et al.~\cite{chen93},
Schulze et al.~\cite{schulze96} and Schwenk et al.~\cite{schwenk03}
predict a large reduction compared to the BCS gap, with a maximum
gap of about 1 MeV.
The microscopic calculations based
on correlated basis function
theory or using quantum Monte Carlo techniques show a reduction of the gap
compared to the BCS result particularly at high densities, where the
maximum is about $2.1$ MeV using AFDMC and $2.4$ MeV with correlated
basis functions. The other
available quantum Monte Carlo
result by Gezerlis and Carlson was performed for smaller
densities because it neglects several contributions from other channels
of the interaction~\cite{gezerlis08}. The recent results provided by
other many--body techniques using Bruckner Hartree-Fock
and new effective interactions by
Cao et al.~\cite{cao06} and Margueron et al.~\cite{margueron08} predict a
superfluid gap closer to the AFDMC result.
Their maximum value of $\Delta$
is about $1.7$ MeV. In addition, the different methods
predict different densities where the gap reaches the maximum value.

\subsection{Pair distribution functions and pairing orbitals}
\label{subsec:distr}

Besides computing energies, the
structure of $^1S_0$ pairing can be investigated by a qualitative study of
pair distribution functions. If the pair energy is
low enough that only the $^1S_0$ or the $^1S_0$
and $^3P_1$ channels are important,
the interaction can be written
as $v_c(r_{ij}) +v_\sigma(r_{ij}) \vec \sigma_i \cdot \vec \sigma_j$.
Even though we keep the full interaction, it is interesting to look
at the two-body distributions that have this form.
The corresponding
pair distribution functions are defined by
\begin{equation}
\label{def:g_c}
g_c(r)=\frac{1}{2\pi r^2 \rho N} \sum_{i<j}
\frac{\langle\psi\vert\delta(r_{ij}-r)\vert\psi\rangle}
{\langle\psi\vert\psi\rangle} \,,
\end{equation}
and
\begin{equation}
\label{def:g_s}
g_\sigma(r)=
\frac{1}{2\pi r^2 \rho N}
\sum_{i<j}\frac{\langle\psi\vert\delta(r_{ij}-r)
\vec\sigma_i\cdot\vec\sigma_j\vert\psi\rangle}
{\langle\psi\vert\psi\rangle} \,,
\end{equation}
where $\rho$ is the density. 
$\rho g_c(r) d^3r$ is the probability of finding a neutron
in an infinitesimal
volume $d^3r$ at a distance $r$ from another neutron, while
$\rho g_\sigma(r) d^3r$
is -3 times the probability of finding a neutron such that the two
are in a singlet state plus the probability of finding a neutron such that
the two are in a triplet state.
In the limit of large $r$, $g_c(r)\to1$, while $g_\sigma(r)\to0$.

Since $\vec\sigma_i\cdot\vec\sigma_j$ is 1 in triplet and -3 in singlet channels,
we can write singlet and triplet pair distribution functions,
$g_{S}(r)$, where $S=0$ for the singlet and $S=1$ for the triplet,
\begin{equation}
g_{0}(r)=\dfrac{1}{4}[g_c(r)-g_\sigma(r)] \,,
\end{equation}
and
\begin{equation}
g_{1}(r)=\dfrac{1}{4}[3g_c(r)+g_\sigma(r)] \,.
\end{equation}

Because
AFDMC, like diffusion Monte Carlo, most easily calculates mixed
estimates
\begin{equation}
\langle O \rangle_M = \frac{\langle \psi|O|\psi_T\rangle}
{\langle \psi|\psi_T\rangle} \,,
\end{equation}
we extrapolate these from the variational values
\begin{equation}
\langle O \rangle_V = \frac{\langle \psi_T|O|\psi_T\rangle}
{\langle \psi_T|\psi_T\rangle}
\end{equation}
as $\langle O\rangle \simeq 2\langle O\rangle_M-\langle O\rangle_V$.

\begin{figure}[ht]
\vspace{1cm}
\begin{center}
\includegraphics[width=8cm]{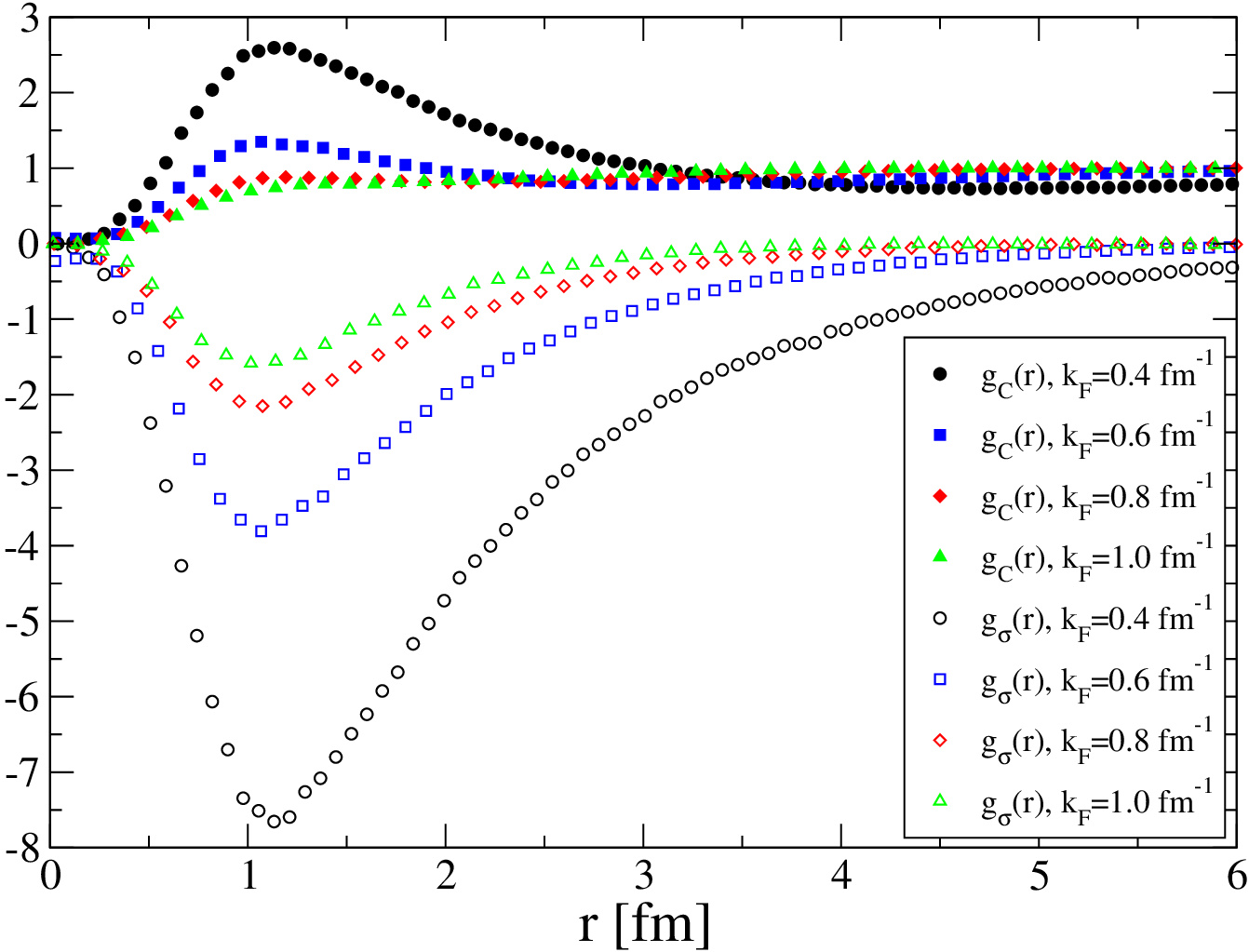}
\vspace{0.5cm}
\caption{(color online) Pair distribution functions $g_c(r)$ and
$g_\sigma(r)$ for neutron matter as defined in the text.
The curves with closed symbols are
the $g_c(r)$, those with open ones indicate $g_\sigma(r)$ corresponding
to different Fermi wave vector. See the text for details.}
\label{fig:gofr}
\end{center}
\end{figure}

The pair distribution functions computed with AFDMC are shown in
Fig.~\ref{fig:gofr}. Closed symbols refer to $g_c(r)$ at various
densities, while open symbols represent $g_\sigma(r)$. The calculations
were performed at different Fermi wave vector; black circles represents
the $g(r)$ at $k_F=0.4$~fm$^{-1}$, blue squares $k_F=0.6$~fm$^{-1}$, red
diamonds $k_F=0.8$~fm$^{-1}$ and green triangles $k_F=1.0$~fm$^{-1}$. As
it can be seen, the strong interaction in the $^1S_0$ channel is
evident in both the $g_c(r)$ and $g_\sigma(r)$
which exhibit a peak at the same distance. The peak value of $g_\sigma(r)$
is about -3 times that of $g_c(r)$, and the peaks
increase as the density is lowered.

\begin{figure}[ht]
\vspace{1cm}
\begin{center}
\includegraphics[width=8cm]{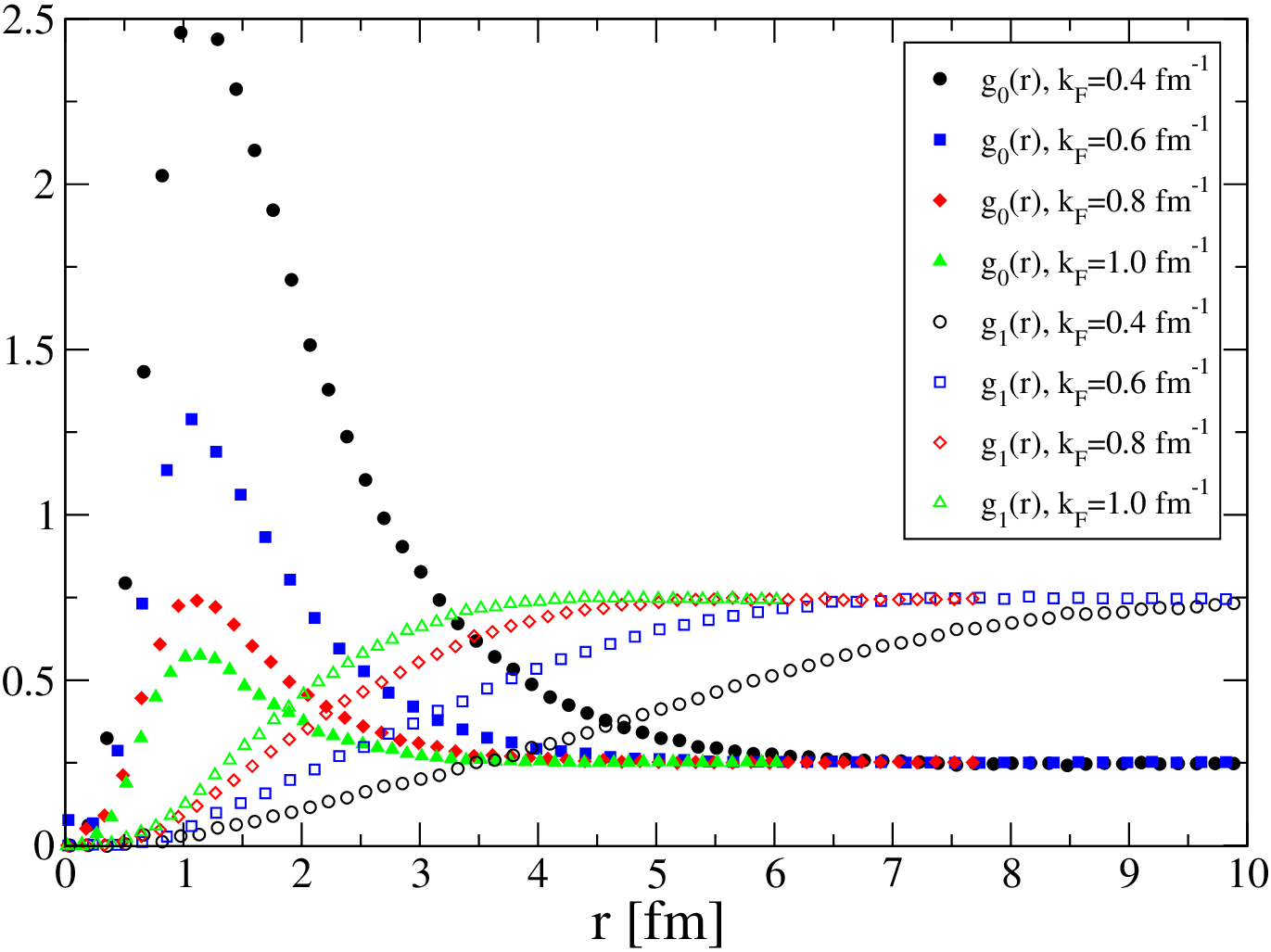}
\vspace{0.5cm}
\caption{(color online) Pair distribution functions $g_{0}(r)$ and
$g_{1}(r)$ as defined in the text.  Closed symbols represent the pair
distribution function projected into the singlet spin channel, while
open ones the triplet spin channel. See the text for details.}
\label{fig:gofr2}
\end{center}
\end{figure}

The strong $^1S_0$ correlation
is more evident using the singlet
and triplet channel distribution functions,
which we show in Fig.~\ref{fig:gofr2}.
Closed symbols represent the
singlet state of the pair, while open ones the triplet state at various Fermi
wave vectors as indicated in the legend. The singlet channel
becomes very strong and dominant when the density decreases.

We can compare these pair distribution functions with
those of a noninteracting Fermi gas,
\begin{equation}
g^{FG}_0(r)=\frac{1}{4}\left[1+l^2(r)\right] \,,
\end{equation}
and
\begin{equation}
g^{FG}_1(r)=\frac{3}{4}\left[1-l^2(r)\right] \,,
\end{equation}
where $l(r)$ is the Slater function defined as
\begin{equation}
l(r)=3\frac{\sin(k_Fr)-k_Fr\cos(k_Fr)}{(k_Fr)^3} \,.
\end{equation}

\begin{figure}
\begin{minipage}[ht]{0.5\textwidth}
\begin{center}
\includegraphics[width=8cm]{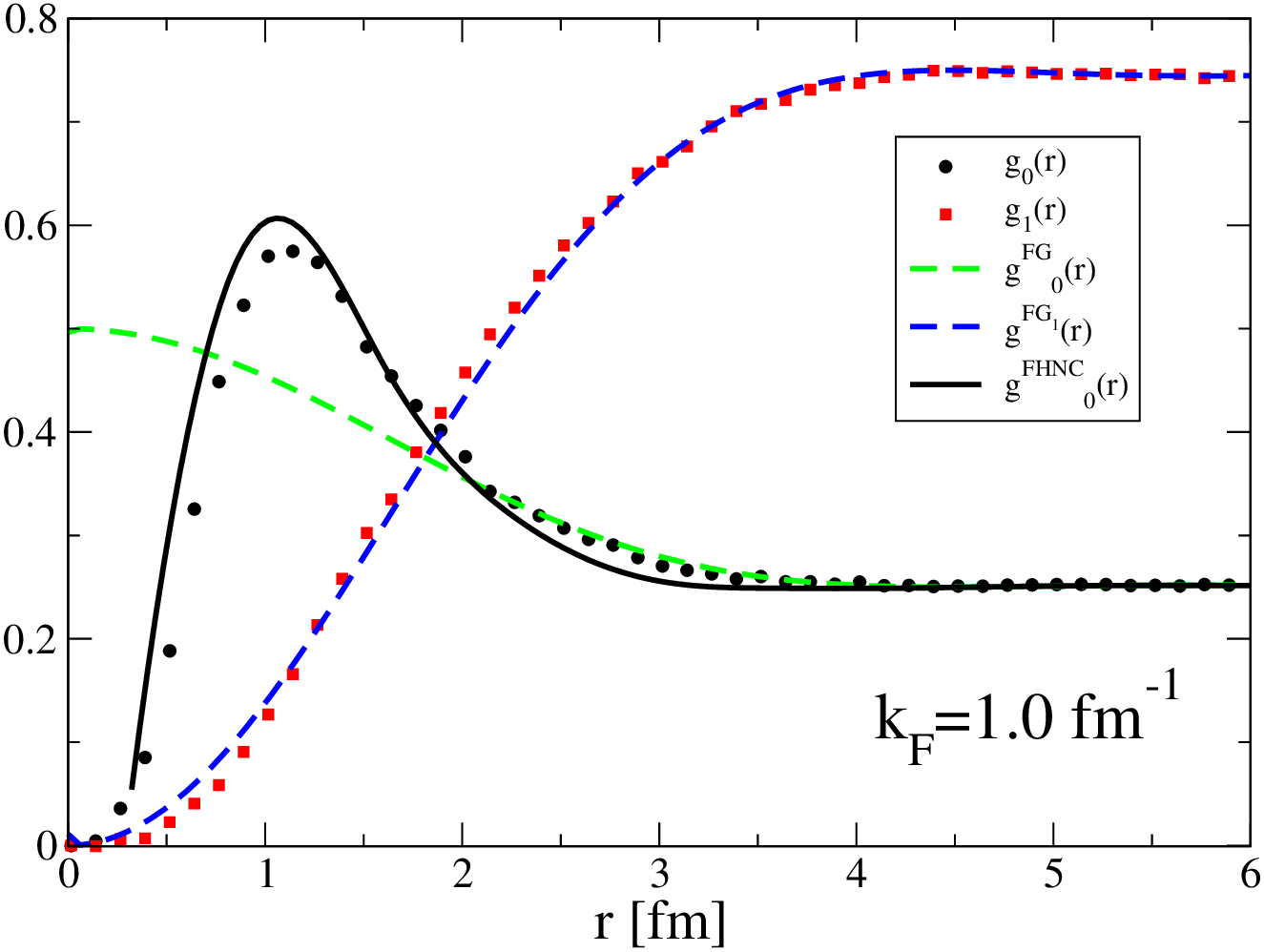}
\vspace{1.0cm}
\end{center}
\end{minipage}
\begin{minipage}[ht]{0.5\textwidth}
\begin{center}
\includegraphics[width=8cm]{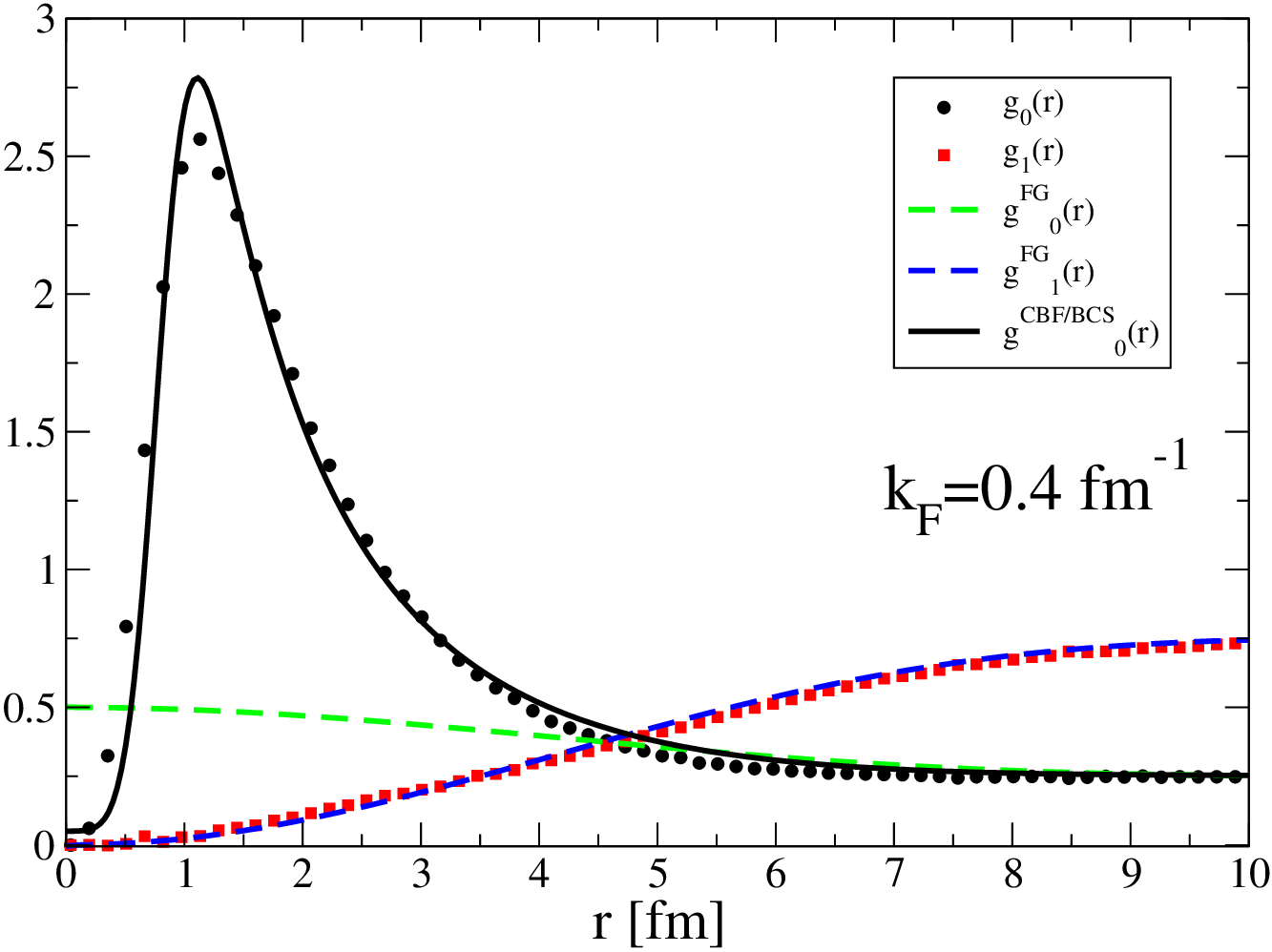}
\vspace{0.5cm}
\end{center}
\end{minipage}
\caption{(color online) Pair distribution functions $g_{0}(r)$ and
$g_{1}(r)$, as defined in the text, at $k_F=1.0$~fm$^{-1}$ (upper
panel) and $k_F=0.4$~fm$^{-1}$ (lower panel). See the text for details.}
\label{fig:gofr3}
\end{figure}

We report in Fig. \ref{fig:gofr3} $g_{0}(r)$ (black circles) and
$g_{1}(r)$ (red squares), and the corresponding $g^{FG}_0(r)$ (green
dashed lines) and $g^{FG}_1(r)$ (blue dashed lines) of the Fermi gas at
Fermi wave vector $k_F=1.0$~fm$^{-1}$ and $k_F=0.4$~fm$^{-1}$.
The triplet pair distribution function does not differ
very much
from the noninteracting case; it does have a small deviation at large
distances for $k_F=1.0$~fm$^{-1}$. This means that quantum correlations, in
this channel, in this density regime, are not too important.
They become relatively more important at higher densities.
The singlet pair distribution function, instead, is completely
different than that of the noninteracting Fermi gas.
However, at $k_F=1.0$~fm$^{-1}$ the
peak of $g_{0}(r)$ is not so very far from the maximum value of $g^{FG}_0(r)$
at the origin, while at $k_F=0.4$~fm$^{-1}$ the strong peak of the singlet
is far from the noninteracting case.
The singlet pair distribution function is also compared with corresponding
variational correlated basis function calculations (black solid lines) using either
Fermi hyper-netted chain (FHNC) approach \cite{pandharipande79} for the normal phase
or CBF/BCS \cite{fabrocini08} for the superfluid phase. It is evident that
the strong peak of the singlet is due to presents of the strong correlations
in the system.

\begin{figure}[ht]
\vspace{1cm}
\begin{center}
\includegraphics[width=8cm]{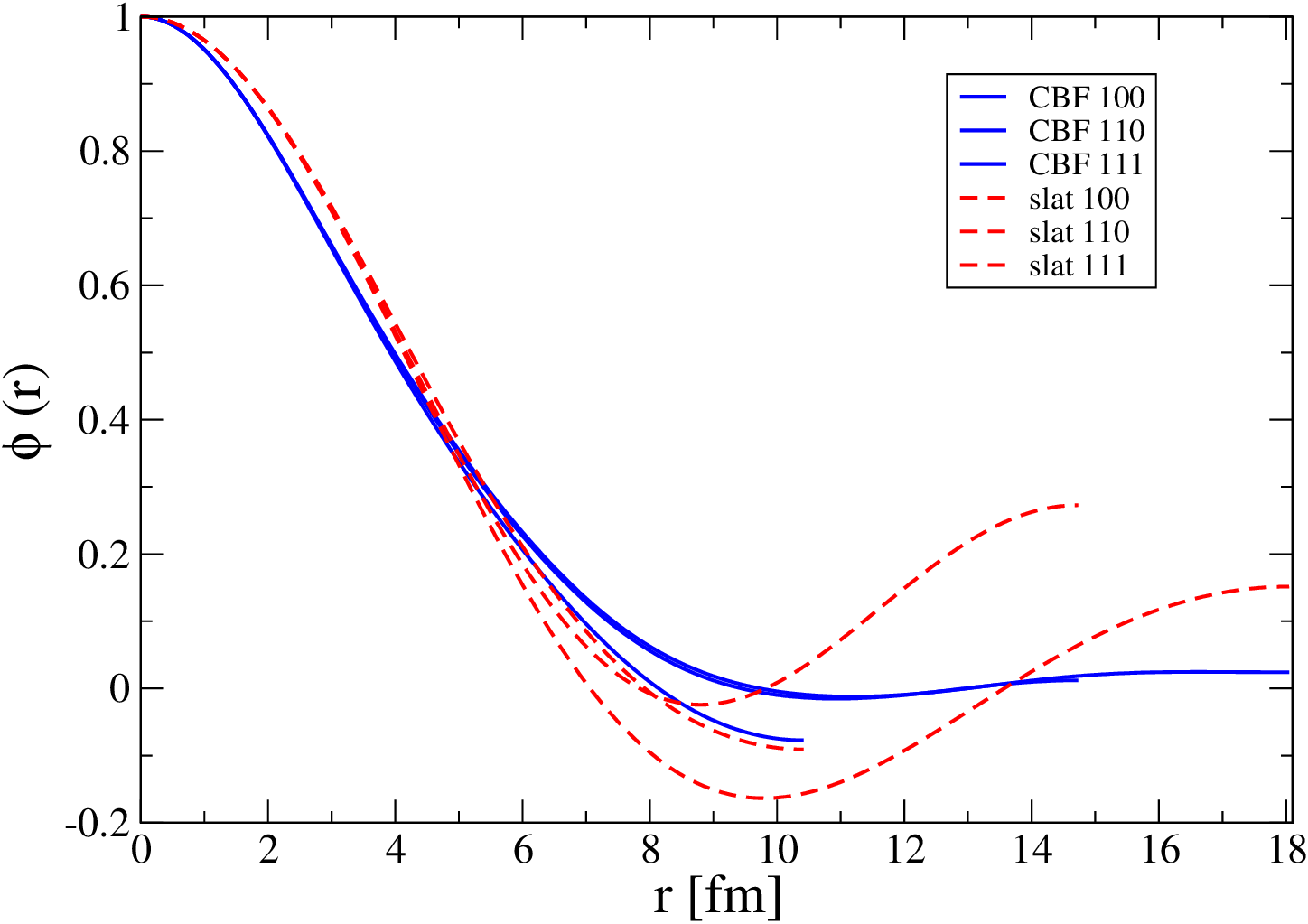}
\vspace{0.5cm}
\caption{(color online) The spatial functions used in the pairing orbitals
at $k_F=0.6$~fm$^{-1}$.  The solid (blue) line is the function
obtained using the correlated basis function (CBF)
coefficients while the dotted (red) line is the simulation cell
Slater function.}
\label{fig:phi}
\end{center}
\end{figure}

We plot in Fig.~\ref{fig:phi} the spatial part
of the pairing function
used in $\Phi_{BCS}$ at $k_F=0.6$~fm$^{-1}$,
along the three spatial directions 100, 110 and 111 obtained by
using the correlated basis function
coefficients. These are compared with the
simulation cell Slater functions $\ell_{\rm cell} =
\frac{2}{N}\sum_{\vec k, k < k_F} e^{i \vec k \cdot \vec r}$.
The functions corresponding to each direction end at $L/2$, $L/\sqrt2$
and $\sqrt{3/4}L$, where $L$ is the side of the simulation cell.

\section{Conclusion}
We have reported a detailed computation of the equation of state
of neutron matter in the
low--density regime where the system is superfluid and neutrons pair
in the $^1S_0$ channel. The superfluid gap was also computed.
The presence of spin--dependent interactions means
that
the wave function must be written as a pfaffian of two--neutrons pairing
orbitals, and the definition and the computation of the pfaffian was also
discussed.

The use of a realistic nuclear Hamiltonian without using any effective
interaction combined with the use of a very accurate projection technique
makes these results benchmark for other methods.
Because of the constraint used to control the
Fermion sign problem, the results could in principle depend on the
importance function. We carefully verified the effect of the wave
function without observing a particular bias due to the fixed--phase
constraint used in the calculations.

We compared the computed equation of state
with other results, and we observed important
deviations that could be due both to the model Hamiltonian and to the methods
used to solve for the ground state.  We found that the $^1S_0$ pairing gap is
only somewhat lower than that predicted by the simple BCS theory
for densities corresponding to $k_F<0.5$~fm$^{-1}$, but the polarization
effects due to the bulk are very important at higher densities where a large
suppression of the maximum value of the gap with respect to the BCS prediction
was found.  In particular, the maximum value
of the gap is a bit larger with respect to other recent calculations,
and much larger than other calculations based on effective interactions.

\begin{acknowledgments}
We thank A.~Gezerlis, J.~Carlson, J.~Margueron and C.~Pethick for
useful discussions.  Calculations were partially performed on the BEN
cluster at ECT$^\star$ in Trento, under a grant for supercomputing
projects, partially on the HPC facility ``WIGLAF'' of the Department
of Physics, University of Trento, and partially on the HPC facility of
SISSA/Democritos in Trieste. 
This work was supported in part by the NSF grant PHY-0757703.
\end{acknowledgments}

\appendix

\section{BCS wave function projected to fixed N}
\label{appendixa}
The original BCS wave function was not an eigenstate of
particle number -- i.e. it explicitly broke gauge symmetry. For
spin-singlet paired fermions with the pairs having total momentum zero,
the BCS form can be written as
\begin{equation}
|{\rm BCS}\rangle \propto \prod_{\vec k}
\left [ u_k + v_k a^+_{\vec k \uparrow} a^+_{-\vec k \downarrow} \right ]
|0\rangle
\end{equation}
where the $a^+_{\vec k s}$ is the fermion creation operator for a
particle in the $\vec k$ wave vector and spin projection $s$ state,
with anticommutation relations
\begin{equation}
\{a_{\vec k s}, a^+_{\vec k' s'}\} = \delta_{\vec k, \vec k'} \delta_{s,s'} \,.
\end{equation}
The $u_k$ and $v_k$ here are functions only of the magnitude, $k = |\vec k|$,
and this spatial symmetry along with the fermion antisymmetry guarantees
only singlet pairs.

For our Monte Carlo calculations, it is simpler to use
the projection of this state onto a fixed number of particles, $N$,
in a periodic simulation cell of side $L$.
We write the antisymmetric position- and spin-projected states as
\begin{eqnarray}
&&
{\cal A} |\vec r_1,s_1,\vec r_2, s_2, ... \vec r_N, s_N\rangle
\nonumber\\
&&
= \frac{1}{N!} \sum_{\rm permutations\ P} (-1)^P 
|P \left ( \vec r_1,s_1,\vec r_2, s_2, ... \vec r_N, s_N \right ) \rangle
\nonumber\\
&&
= \frac{1}{\sqrt{N!}}
\psi^+_{s_1}(\vec r_1)
\psi^+_{s_2}(\vec r_2) ...
\psi^+_{s_N}(\vec r_N) |0\rangle \,,
\label{eq:antisym}
\end{eqnarray}
where $P$ represents the permutation of the particle labels, and $(-1)^P$
is 1 (-1) for even (odd) permutations.
The position and momentum creation operators are related by
\begin{equation}
a^+_{\vec k s} = \frac{1}{L^{3/2}}
\int_{-L/2}^{L/2}dx
\int_{-L/2}^{L/2}dy
\int_{-L/2}^{L/2}dz
e^{i \vec k \cdot \vec r} \psi^+_s(\vec r) \,.
\end{equation}

The standard BCS state is usually normalized by
choosing $|u_k|^2 + |v_k|^2 = 1$ . Since we will be projecting out the
part with $N$ particles, even if we start with a normalized
state, the projected part will no longer be normalized.
There is then no advantage to taking a normalized state, and instead
we divide by each of the $u_k$. If one or more are
zero, it simply means that we should drop the $1$ term for that $k$
value since it is always filled. We therefore take
\begin{equation}
|{\rm BCS}\rangle = \prod_{\vec k}
\left [ 1 + \frac{v_k}{u_k}
a^+_{\vec k \uparrow} a^+_{-\vec k \downarrow} \right ] |0\rangle \,.
\end{equation}

The particle-projected BCS wave function is then
\begin{widetext}
\begin{equation}
\label{eq.bcsprojected}
\Psi_{BCS} (\vec R,S) = \langle \vec R,S|{\rm BCS}\rangle = \frac{1}{\sqrt{N!}} \langle 0 |
\psi_{s_N}(\vec r_N)
\psi_{s_{N-1}}(\vec r_{N-1})
...
\psi_{s_1}(\vec r_1) 
\prod_{\vec k}
\left [ 1 + \frac{v_k}{u_k} 
a^+_{\vec k \uparrow} a^+_{-\vec k \downarrow}
\right ] |0\rangle \,.
\end{equation}
\end{widetext}

This is readily evaluated using Wick's theorem\cite{wick50} to change from the
given order to the normal order. Contracting $\psi_s(\vec r)$
and $a^+_{\vec k s}$ gives
\begin{equation}
\contraction{}{\psi}{{}_s(\vec r)}{a}
\psi_s(\vec r)a_{\vec k s'}^+ =
L^{-3/2} e^{i \vec k \cdot \vec r} \delta_{ss'} \,.
\end{equation}
From Eq. \ref{eq.bcsprojected}, we see that either both
$a^+_{\vec k \uparrow}$ and $a^+_{-\vec k\downarrow}$ in a pair
or neither
must be contracted with $\psi_{s}(\vec r)$ operators to give a nonzero
result. One particular contraction occurs when
$\psi_{s_1}(\vec r_1)$ and 
$\psi_{s_2}(\vec r_2)$ contract with such a pair in $\vec k_1$,
$\psi_{s_3}(\vec r_3)$ and 
$\psi_{s_4}(\vec r_4)$ contract with another pair in $\vec k_2$, etc.
This gives a term
\begin{eqnarray}
&&
\frac{v_{k_1}}{u_{k_1}} e^{i \vec k_1 \cdot (\vec r_1 - \vec r_2)}
\langle s_1 s_2 |\uparrow \downarrow\rangle
\frac{v_{k_2}}{u_{k_2}} e^{i \vec k_2 \cdot (\vec r_3 - \vec r_4)}
\langle s_3 s_4 |\uparrow \downarrow\rangle
...
\nonumber\\
&&
\frac{v_{k_{N/2}}}{u_{k_{N/2}}}
e^{i \vec k_{N/2} \cdot (\vec r_{N-1} - \vec r_N)}
\langle s_{N-1} s_N |\uparrow \downarrow\rangle \,,
\end{eqnarray}
where we drop an unimportant overall normalization factor.
Choosing different $\vec k$ terms to contract with corresponds to
summing over all values of the $\vec k_1$, $\vec k_2$, etc. with the
constraint that no two of the $\vec k_n$  values should be equal
(anticommutating two pairs of operators does not change the sign).
Choosing other contractions completely antisymmetrizes this form, and we can
then include all terms in the $\vec k$ sums since these cancel when
antisymmetrized. The result is
\begin{equation}
\label{eq.bcsconfig}
\Phi_{BCS}={\cal A} \left[\phi(\vec r_1,s_1,\vec r_2,s_2)\dots
\phi(\vec r_{N-1},s_{N-1},\vec r_N, s_N)\right] \,.
\end{equation}
where, for spin-singlet zero-momentum pairs,
\begin{equation}
\phi(\vec r_1,s_1,\vec r_2,s_2) =
\sum_{\vec k} \frac{v_k}{u_k}e^{i \vec k \cdot (\vec r_1-\vec r_2)}
\left [
\langle s_1 s_2 |\uparrow \downarrow\rangle
\rangle \right ] \,.
\end{equation}
Since the many-body antisymmetrizer will interchange the particles in
$\phi$, we usually explicitly antisymmetrize $\phi$. We then get, up
to an unimportant normalization,
\begin{equation}
\phi(\vec r_1,s_1,\vec r_2,s_2) =
\sum_{\vec k} \frac{v_k}{u_k}e^{i \vec k \cdot (\vec r_1-\vec r_2)}
\left [
\langle s_1 s_2 |\uparrow \downarrow\rangle-\langle s_1 s_2 |\downarrow\uparrow
\rangle \right ] \,,
\end{equation}
which explicitly demonstrates the singlet pairing. For a very large
simulation cell, the spatial function would be spherically symmetric and
therefore an $S$ state. For the typical sizes of our
simulation cells, the function has the symmetry of the cube as seen
in Fig. \ref{fig:phi}.
Other possible fully paired
states have different $\phi(\vec r_1,s_1,\vec r_2,s_2)$,
but still have the general form of Eq. \ref{eq.bcsconfig}.

Often we want to investigate systems which are not fully paired.
Obviously, if we have an odd number of particles, at least one must
be unpaired. We include unpaired
particles in specific states by multiplying the $|{\rm BCS}\rangle$
state by a product of creation operators (or linear combinations of
creation operators) for those states.
The only change to the particle number projection described above
is that these creation operators must be contracted with one of the
$\psi_s(\vec r)$ or the result will be zero. For $n$ pairs and $o$
occupied single particle states, we have
\begin{equation}
\Phi_{BCS} = {\cal A} \left [
\phi_{12}\phi_{34} ...\phi_{2n-1,2n} \psi_1(2n+1) ...\psi_o(N) \right ]
\end{equation}
which is Eq. \ref{2n+o}.

\section{Pfaffian calculations}
\label{appendixb}
Here we give some details on how to calculate the pfaffian. Proofs
of the statements are
given in ref. ~\cite{bajdich08}.
The pfaffian of a skew-symmetric
matrix has the following three properties:
\begin{itemize}
\item[a.]
Multiplying a row and the corresponding column by a constant is
equivalent to multiplying the pfaffian by a constant.
\item[b.]
Interchanging two different rows and the corresponding columns changes
the sign of the Pfaffian.
\item[c.]
A multiple of a row and corresponding column added to another row
and corresponding column does not change the value of the pfaffian.
\end{itemize}
In addition, the matrix must have even rank for the pfaffian to be
nonzero.
Using these properties, it is straightforward to use, for example,
Gauss elimination to reduce the skew-symmetric
matrix to a block diagonal form with $2\times 2$ blocks, whose pfaffian
is just the product of the nonzero elements in the first superdiagonal.
A Fortran fragment showing the algorithm without pivoting 
for a complex matrix $a$ of even rank n, is
\begin{verbatim}
   p=(1.0,0.0)
   do i=1,n,2
      do j=i+2,n
         fac=-a(i,j)/a(i,i+1)
         a(i+1:n,j)=a(i+1:n,j)+fac*a(i+1:n,i+1)
         a(j,i+1:n)=a(j,i+1:n)+fac*a(i+1,i+1:n)
      enddo
      p=p*a(i,i+1)
   enddo
\end{verbatim}
As in standard Gauss elimination, we search the current row for a
large pivot element, and pivot using property b to bring this onto
the superdiagonal so that we don't divide by small numbers a(i,i+1).

At the same time, we calculate the inverse of the matrix.

When one particle changes position or spin (or for calculation of
one-body properties like the gradient, kinetic energy or expectation
of a spin operator) the skew-symmetric matrix $A$ has one row and
the corresponding column changed. Writing the matrix $B$ to be equal to
$A$ except for the row $k$ with new elements $B_{kj}$ and the corresponding
column elements, Cayley showed~\cite{cayley1849}
\begin{equation}
{\rm Pf}[B] = {\rm Pf}[A] \sum_j B_{kj} A^{-1}_{jk} \,.
\end{equation}

For efficient algorithms with spin-dependent potentials
we want to be able to change two particles.
A straightforward implementation would first change one row of the
matrix as above and calculate the new pfaffian, and
update the inverse. Then change the corresponding column to obtain
the skew-symmetric matrix and its inverse (its determinant is the
square of the pfaffian obtained before). This will require $O(N^2)$
operations. For each of the $N$ second particles we will require
$O(N)$ operations to calculate the new pfaffian if the first column
is different for each pair. Unfortunately the
result is $O(N^4)$ to calculate pairwise potentials.

However, for our case, the operation needed on a column or row is independent
of the other column or row (except for the common element). We can
therefore imagine doing a single update for particle 1 and using this for all
the terms where the pair contains particle 1. The common element does
not require an update and can be done separately.

It is most efficient to write this as a set of matrix multiplies.
We define the new column $j$ of the matrix to
be $C_{ij}$, corresponding to a spin or derivative operator on particle
$j$. Defining
\begin{align}
P_{ij} &= \sum_k A^{-1}_{ik} C_{kj} \,,
\nonumber\\
G_{ij} &= \sum_{mk} C^T_{im} A^{-1}_{mk} C_{kj}
= \sum_m C_{mi} P_{mj} = -G_{ji} \,,
\end{align}
we find that the ratio of the new to old pfaffians with the two rows and
columns denoted by $i$ and $j$ changed is
\begin{equation}
\frac{{\rm Pf(new})}{{\rm Pf(old})}
= A^{-1}_{ji} [ A^{new}_{ij} +G_{ij} ]
+ P_{ii} P_{jj} -P_{ij}P_{ji} \,,
\end{equation}
where $A^{new}$ is the $A$-matrix with new rows and columns.


\begin{thebibliography}{54}
\expandafter\ifx\csname natexlab\endcsname\relax\def\natexlab#1{#1}\fi
\expandafter\ifx\csname bibnamefont\endcsname\relax
  \def\bibnamefont#1{#1}\fi
\expandafter\ifx\csname bibfnamefont\endcsname\relax
  \def\bibfnamefont#1{#1}\fi
\expandafter\ifx\csname citenamefont\endcsname\relax
  \def\citenamefont#1{#1}\fi
\expandafter\ifx\csname url\endcsname\relax
  \def\url#1{\texttt{#1}}\fi
\expandafter\ifx\csname urlprefix\endcsname\relax\def\urlprefix{URL }\fi
\providecommand{\bibinfo}[2]{#2}
\providecommand{\eprint}[2][]{\url{#2}}

\bibitem[{\citenamefont{Pethick and Ravenhall}(1995)}]{pethick95a}
\bibinfo{author}{\bibfnamefont{C.}~\bibnamefont{Pethick}} \bibnamefont{and}
  \bibinfo{author}{\bibfnamefont{D.}~\bibnamefont{Ravenhall}},
  \bibinfo{journal}{Annu. Rev. Part. Sci.} \textbf{\bibinfo{volume}{45}},
  \bibinfo{pages}{429} (\bibinfo{year}{1995}).

\bibitem[{\citenamefont{Heiselberg and Pandharipande}(2000)}]{heiselberg00}
\bibinfo{author}{\bibfnamefont{H.}~\bibnamefont{Heiselberg}} \bibnamefont{and}
  \bibinfo{author}{\bibfnamefont{V.}~\bibnamefont{Pandharipande}},
  \bibinfo{journal}{Annu. Rev. Nucl. Part. Sci.} \textbf{\bibinfo{volume}{50}},
  \bibinfo{pages}{481} (\bibinfo{year}{2000}).

\bibitem[{\citenamefont{Stoks et~al.}(1993)\citenamefont{Stoks, Timmermans, and
  de~Swart}}]{stoks93}
\bibinfo{author}{\bibfnamefont{V.}~\bibnamefont{Stoks}},
  \bibinfo{author}{\bibfnamefont{R.}~\bibnamefont{Timmermans}},
  \bibnamefont{and} \bibinfo{author}{\bibfnamefont{J.~J.}
  \bibnamefont{de~Swart}}, \bibinfo{journal}{Phys. Rev. C}
  \textbf{\bibinfo{volume}{47}}, \bibinfo{pages}{512} (\bibinfo{year}{1993}).

\bibitem[{\citenamefont{Gandolfi
  et~al.}(2008{\natexlab{a}})\citenamefont{Gandolfi, Illarionov, Fantoni,
  Pederiva, and Schmidt}}]{gandolfi08b}
\bibinfo{author}{\bibfnamefont{S.}~\bibnamefont{Gandolfi}},
  \bibinfo{author}{\bibfnamefont{A.~Y.} \bibnamefont{Illarionov}},
  \bibinfo{author}{\bibfnamefont{S.}~\bibnamefont{Fantoni}},
  \bibinfo{author}{\bibfnamefont{F.}~\bibnamefont{Pederiva}}, \bibnamefont{and}
  \bibinfo{author}{\bibfnamefont{K.~E.} \bibnamefont{Schmidt}},
  \bibinfo{journal}{Phys. Rev. Lett.} \textbf{\bibinfo{volume}{101}},
  \bibinfo{pages}{132501} (\bibinfo{year}{2008}{\natexlab{a}}).

\bibitem[{\citenamefont{Schmidt and Fantoni}(1999)}]{schmidt99}
\bibinfo{author}{\bibfnamefont{K.~E.} \bibnamefont{Schmidt}} \bibnamefont{and}
  \bibinfo{author}{\bibfnamefont{S.}~\bibnamefont{Fantoni}},
  \bibinfo{journal}{Phys. Lett. B} \textbf{\bibinfo{volume}{446}},
  \bibinfo{pages}{99} (\bibinfo{year}{1999}).

\bibitem[{\citenamefont{Anderson}(1975)}]{anderson75}
\bibinfo{author}{\bibfnamefont{J.~B.} \bibnamefont{Anderson}},
  \bibinfo{journal}{J. Chem. Phys.} \textbf{\bibinfo{volume}{63}},
  \bibinfo{pages}{1499} (\bibinfo{year}{1975}).

\bibitem[{\citenamefont{Carlson}(1987)}]{carlson87}
\bibinfo{author}{\bibfnamefont{J.}~\bibnamefont{Carlson}},
  \bibinfo{journal}{Phys. Rev. C} \textbf{\bibinfo{volume}{36}},
  \bibinfo{pages}{2026} (\bibinfo{year}{1987}).

\bibitem[{\citenamefont{Pieper}(2005)}]{pieper05}
\bibinfo{author}{\bibfnamefont{S.~C.} \bibnamefont{Pieper}},
  \bibinfo{journal}{Nucl. Phys. A} \textbf{\bibinfo{volume}{751}},
  \bibinfo{pages}{516} (\bibinfo{year}{2005}).

\bibitem[{\citenamefont{Gandolfi
  et~al.}(2007{\natexlab{a}})\citenamefont{Gandolfi, Pederiva, Fantoni, and
  Schmidt}}]{gandolfi07b}
\bibinfo{author}{\bibfnamefont{S.}~\bibnamefont{Gandolfi}},
  \bibinfo{author}{\bibfnamefont{F.}~\bibnamefont{Pederiva}},
  \bibinfo{author}{\bibfnamefont{S.}~\bibnamefont{Fantoni}}, \bibnamefont{and}
  \bibinfo{author}{\bibfnamefont{K.~E.} \bibnamefont{Schmidt}},
  \bibinfo{journal}{Phys. Rev. Lett.} \textbf{\bibinfo{volume}{99}},
  \bibinfo{pages}{022507} (\bibinfo{year}{2007}{\natexlab{a}}).

\bibitem[{\citenamefont{Gandolfi et~al.}(2006)\citenamefont{Gandolfi, Pederiva,
  Fantoni, and Schmidt}}]{gandolfi06}
\bibinfo{author}{\bibfnamefont{S.}~\bibnamefont{Gandolfi}},
  \bibinfo{author}{\bibfnamefont{F.}~\bibnamefont{Pederiva}},
  \bibinfo{author}{\bibfnamefont{S.}~\bibnamefont{Fantoni}}, \bibnamefont{and}
  \bibinfo{author}{\bibfnamefont{K.~E.} \bibnamefont{Schmidt}},
  \bibinfo{journal}{Phys. Rev. C} \textbf{\bibinfo{volume}{73}},
  \bibinfo{pages}{044304} (\bibinfo{year}{2006}).

\bibitem[{\citenamefont{Gandolfi
  et~al.}(2008{\natexlab{b}})\citenamefont{Gandolfi, Pederiva, and
  a~Beccara}}]{gandolfi08}
\bibinfo{author}{\bibfnamefont{S.}~\bibnamefont{Gandolfi}},
  \bibinfo{author}{\bibfnamefont{F.}~\bibnamefont{Pederiva}}, \bibnamefont{and}
  \bibinfo{author}{\bibfnamefont{S.}~\bibnamefont{a~Beccara}},
  \bibinfo{journal}{Eur. Phys. J. A} \textbf{\bibinfo{volume}{35}},
  \bibinfo{pages}{207} (\bibinfo{year}{2008}{\natexlab{b}}).

\bibitem[{\citenamefont{Sarsa et~al.}(2003)\citenamefont{Sarsa, Fantoni,
  Schmidt, and Pederiva}}]{sarsa03}
\bibinfo{author}{\bibfnamefont{A.}~\bibnamefont{Sarsa}},
  \bibinfo{author}{\bibfnamefont{S.}~\bibnamefont{Fantoni}},
  \bibinfo{author}{\bibfnamefont{K.~E.} \bibnamefont{Schmidt}},
  \bibnamefont{and} \bibinfo{author}{\bibfnamefont{F.}~\bibnamefont{Pederiva}},
  \bibinfo{journal}{Phys. Rev. C} \textbf{\bibinfo{volume}{68}},
  \bibinfo{pages}{024308} (\bibinfo{year}{2003}).

\bibitem[{\citenamefont{Gandolfi et~al.}(2009)\citenamefont{Gandolfi,
  Illarionov, Schmidt, Pederiva, and Fantoni}}]{gandolfi09}
\bibinfo{author}{\bibfnamefont{S.}~\bibnamefont{Gandolfi}},
  \bibinfo{author}{\bibfnamefont{A.~Y.} \bibnamefont{Illarionov}},
  \bibinfo{author}{\bibfnamefont{K.~E.} \bibnamefont{Schmidt}},
  \bibinfo{author}{\bibfnamefont{F.}~\bibnamefont{Pederiva}}, \bibnamefont{and}
  \bibinfo{author}{\bibfnamefont{S.}~\bibnamefont{Fantoni}},
  \bibinfo{journal}{Phys. Rev. C.} \textbf{\bibinfo{volume}{79}},
  \bibinfo{pages}{054005} (\bibinfo{year}{2009}).

\bibitem[{\citenamefont{Gandolfi
  et~al.}(2007{\natexlab{b}})\citenamefont{Gandolfi, Pederiva, Fantoni, and
  Schmidt}}]{gandolfi07}
\bibinfo{author}{\bibfnamefont{S.}~\bibnamefont{Gandolfi}},
  \bibinfo{author}{\bibfnamefont{F.}~\bibnamefont{Pederiva}},
  \bibinfo{author}{\bibfnamefont{S.}~\bibnamefont{Fantoni}}, \bibnamefont{and}
  \bibinfo{author}{\bibfnamefont{K.~E.} \bibnamefont{Schmidt}},
  \bibinfo{journal}{Phys. Rev. Lett.} \textbf{\bibinfo{volume}{98}},
  \bibinfo{pages}{102503} (\bibinfo{year}{2007}{\natexlab{b}}).

\bibitem[{\citenamefont{Friedman and Pandharipande}(1981)}]{friedman81}
\bibinfo{author}{\bibfnamefont{B.}~\bibnamefont{Friedman}} \bibnamefont{and}
  \bibinfo{author}{\bibfnamefont{V.}~\bibnamefont{Pandharipande}},
  \bibinfo{journal}{Nucl. Phys. A} \textbf{\bibinfo{volume}{361}},
  \bibinfo{pages}{502} (\bibinfo{year}{1981}).

\bibitem[{\citenamefont{Akmal et~al.}(1998)\citenamefont{Akmal, Pandharipande,
  and Ravenhall}}]{akmal98}
\bibinfo{author}{\bibfnamefont{A.}~\bibnamefont{Akmal}},
  \bibinfo{author}{\bibfnamefont{V.~R.} \bibnamefont{Pandharipande}},
  \bibnamefont{and} \bibinfo{author}{\bibfnamefont{D.~G.}
  \bibnamefont{Ravenhall}}, \bibinfo{journal}{Phys. Rev. C}
  \textbf{\bibinfo{volume}{58}}, \bibinfo{pages}{1804} (\bibinfo{year}{1998}).

\bibitem[{\citenamefont{Carlson
  et~al.}(2003{\natexlab{a}})\citenamefont{Carlson, Morales, Pandharipande, and
  Ravenhall}}]{carlson03}
\bibinfo{author}{\bibfnamefont{J.}~\bibnamefont{Carlson}},
  \bibinfo{author}{\bibfnamefont{J.}~\bibnamefont{Morales}},
  \bibinfo{author}{\bibfnamefont{V.~R.} \bibnamefont{Pandharipande}},
  \bibnamefont{and} \bibinfo{author}{\bibfnamefont{D.~G.}
  \bibnamefont{Ravenhall}}, \bibinfo{journal}{Phys. Rev. C}
  \textbf{\bibinfo{volume}{68}}, \bibinfo{pages}{025802}
  (\bibinfo{year}{2003}{\natexlab{a}}).

\bibitem[{\citenamefont{Schwenk and Pethick}(2005)}]{schwenk05}
\bibinfo{author}{\bibfnamefont{A.}~\bibnamefont{Schwenk}} \bibnamefont{and}
  \bibinfo{author}{\bibfnamefont{C.~J.} \bibnamefont{Pethick}},
  \bibinfo{journal}{Phys. Rev. Lett.} \textbf{\bibinfo{volume}{95}},
  \bibinfo{pages}{160401} (\bibinfo{year}{2005}).

\bibitem[{\citenamefont{Borasoy et~al.}(2008)\citenamefont{Borasoy, Epelbaum,
  Krebs, Lee, and Meissner}}]{borasoy08}
\bibinfo{author}{\bibfnamefont{B.}~\bibnamefont{Borasoy}},
  \bibinfo{author}{\bibfnamefont{E.}~\bibnamefont{Epelbaum}},
  \bibinfo{author}{\bibfnamefont{H.}~\bibnamefont{Krebs}},
  \bibinfo{author}{\bibfnamefont{D.}~\bibnamefont{Lee}}, \bibnamefont{and}
  \bibinfo{author}{\bibfnamefont{U.-G.} \bibnamefont{Meissner}},
  \bibinfo{journal}{Eur. Phys. J. A} \textbf{\bibinfo{volume}{35}},
  \bibinfo{pages}{357} (\bibinfo{year}{2008}).

\bibitem[{\citenamefont{Lee}(2009)}]{lee09}
\bibinfo{author}{\bibfnamefont{D.}~\bibnamefont{Lee}}, \bibinfo{journal}{Prog.
  Part. Nucl. Phys.} \textbf{\bibinfo{volume}{63}}, \bibinfo{pages}{117}
  (\bibinfo{year}{2009}).

\bibitem[{\citenamefont{Epelbaum et~al.}(2009)\citenamefont{Epelbaum, Krebs,
  Lee, and Meissner}}]{epelbaum09}
\bibinfo{author}{\bibfnamefont{E.}~\bibnamefont{Epelbaum}},
  \bibinfo{author}{\bibfnamefont{H.}~\bibnamefont{Krebs}},
  \bibinfo{author}{\bibfnamefont{D.}~\bibnamefont{Lee}}, \bibnamefont{and}
  \bibinfo{author}{\bibfnamefont{U.-G.} \bibnamefont{Meissner}},
  \bibinfo{journal}{Eur. Phys. J. A} \textbf{\bibinfo{volume}{40}},
  \bibinfo{pages}{199} (\bibinfo{year}{2009}).

\bibitem[{\citenamefont{Gezerlis and Carlson}(2008{\natexlab{a}})}]{gezerlis08}
\bibinfo{author}{\bibfnamefont{A.}~\bibnamefont{Gezerlis}} \bibnamefont{and}
  \bibinfo{author}{\bibfnamefont{J.}~\bibnamefont{Carlson}},
  \bibinfo{journal}{Phys. Rev. C} \textbf{\bibinfo{volume}{77}},
  \bibinfo{pages}{032801(R)} (\bibinfo{year}{2008}{\natexlab{a}}).

\bibitem[{\citenamefont{Abe and Seki}(2009{\natexlab{a}})}]{abe09}
\bibinfo{author}{\bibfnamefont{T.}~\bibnamefont{Abe}} \bibnamefont{and}
  \bibinfo{author}{\bibfnamefont{R.}~\bibnamefont{Seki}},
  \bibinfo{journal}{Phys. Rev. C} \textbf{\bibinfo{volume}{79}},
  \bibinfo{pages}{054002} (\bibinfo{year}{2009}{\natexlab{a}}).

\bibitem[{\citenamefont{Abe and Seki}(2009{\natexlab{b}})}]{abe09b}
\bibinfo{author}{\bibfnamefont{T.}~\bibnamefont{Abe}} \bibnamefont{and}
  \bibinfo{author}{\bibfnamefont{R.}~\bibnamefont{Seki}},
  \bibinfo{journal}{Phys. Rev. C} \textbf{\bibinfo{volume}{79}},
  \bibinfo{pages}{054003} (\bibinfo{year}{2009}{\natexlab{b}}).

\bibitem[{\citenamefont{Bardeen et~al.}(1957)\citenamefont{Bardeen, Cooper, and
  Schrieffer}}]{bardeen57}
\bibinfo{author}{\bibfnamefont{J.}~\bibnamefont{Bardeen}},
  \bibinfo{author}{\bibfnamefont{L.~N.} \bibnamefont{Cooper}},
  \bibnamefont{and} \bibinfo{author}{\bibfnamefont{J.~R.}
  \bibnamefont{Schrieffer}}, \bibinfo{journal}{Phys. Rev.}
  \textbf{\bibinfo{volume}{108}}, \bibinfo{pages}{1175} (\bibinfo{year}{1957}).

\bibitem[{\citenamefont{Wiringa et~al.}(1995)\citenamefont{Wiringa, Stoks, and
  Schiavilla}}]{wiringa95}
\bibinfo{author}{\bibfnamefont{R.~B.} \bibnamefont{Wiringa}},
  \bibinfo{author}{\bibfnamefont{V.~G.~J.} \bibnamefont{Stoks}},
  \bibnamefont{and}
  \bibinfo{author}{\bibfnamefont{R.}~\bibnamefont{Schiavilla}},
  \bibinfo{journal}{Phys. Rev. C} \textbf{\bibinfo{volume}{51}},
  \bibinfo{pages}{38} (\bibinfo{year}{1995}).

\bibitem[{\citenamefont{Wiringa and Pieper}(2002)}]{wiringa02}
\bibinfo{author}{\bibfnamefont{R.~B.} \bibnamefont{Wiringa}} \bibnamefont{and}
  \bibinfo{author}{\bibfnamefont{S.~C.} \bibnamefont{Pieper}},
  \bibinfo{journal}{Phys. Rev. Lett.} \textbf{\bibinfo{volume}{89}},
  \bibinfo{pages}{182501} (\bibinfo{year}{2002}).

\bibitem[{\citenamefont{Pudliner et~al.}(1997)\citenamefont{Pudliner,
  Pandharipande, Carlson, Pieper, and Wiringa}}]{pudliner97}
\bibinfo{author}{\bibfnamefont{B.~S.} \bibnamefont{Pudliner}},
  \bibinfo{author}{\bibfnamefont{V.~R.} \bibnamefont{Pandharipande}},
  \bibinfo{author}{\bibfnamefont{J.}~\bibnamefont{Carlson}},
  \bibinfo{author}{\bibfnamefont{S.~C.} \bibnamefont{Pieper}},
  \bibnamefont{and} \bibinfo{author}{\bibfnamefont{R.~B.}
  \bibnamefont{Wiringa}}, \bibinfo{journal}{Phys. Rev. C}
  \textbf{\bibinfo{volume}{56}}, \bibinfo{pages}{1720} (\bibinfo{year}{1997}).

\bibitem[{\citenamefont{Pudliner et~al.}(1995)\citenamefont{Pudliner,
  Pandharipande, Carlson, and Wiringa}}]{pudliner95}
\bibinfo{author}{\bibfnamefont{B.~S.} \bibnamefont{Pudliner}},
  \bibinfo{author}{\bibfnamefont{V.~R.} \bibnamefont{Pandharipande}},
  \bibinfo{author}{\bibfnamefont{J.}~\bibnamefont{Carlson}}, \bibnamefont{and}
  \bibinfo{author}{\bibfnamefont{R.~B.} \bibnamefont{Wiringa}},
  \bibinfo{journal}{Phys. Rev. Lett.} \textbf{\bibinfo{volume}{74}},
  \bibinfo{pages}{4396} (\bibinfo{year}{1995}).

\bibitem[{\citenamefont{Fujita and Miyazawa}(1957)}]{fujita57}
\bibinfo{author}{\bibfnamefont{J.}~\bibnamefont{Fujita}} \bibnamefont{and}
  \bibinfo{author}{\bibfnamefont{H.}~\bibnamefont{Miyazawa}},
  \bibinfo{journal}{Prog. Theor. Phys.} \textbf{\bibinfo{volume}{17}},
  \bibinfo{pages}{360} (\bibinfo{year}{1957}).

\bibitem[{\citenamefont{Pieper et~al.}(2001)\citenamefont{Pieper,
  Pandharipande, Wiringa, and Carlson}}]{pieper01}
\bibinfo{author}{\bibfnamefont{S.~C.} \bibnamefont{Pieper}},
  \bibinfo{author}{\bibfnamefont{V.~R.} \bibnamefont{Pandharipande}},
  \bibinfo{author}{\bibfnamefont{R.~B.} \bibnamefont{Wiringa}},
  \bibnamefont{and} \bibinfo{author}{\bibfnamefont{J.}~\bibnamefont{Carlson}},
  \bibinfo{journal}{Phys. Rev. C} \textbf{\bibinfo{volume}{64}},
  \bibinfo{pages}{014001} (\bibinfo{year}{2001}).

\bibitem[{\citenamefont{Gandolfi}(2007)}]{gandolfi07c}
\bibinfo{author}{\bibfnamefont{S.}~\bibnamefont{Gandolfi}}, Ph.D. thesis,
  \bibinfo{school}{University of Trento, Italy} (\bibinfo{year}{2007}),
  \eprint{arXiv:0712.1364 [nucl-th]}.

\bibitem[{\citenamefont{Guardiola}(1998)}]{guardiola98}
\bibinfo{author}{\bibfnamefont{R.}~\bibnamefont{Guardiola}}, in
  \emph{\bibinfo{booktitle}{Microscopic Quantum Many-Body Theories and Their
  Applications, Proceedings of a European Summer School Held at Valencia,
  Spain, 8-19 September 1997}}, edited by
  \bibinfo{editor}{\bibfnamefont{J.}~\bibnamefont{Navarro}} \bibnamefont{and}
  \bibinfo{editor}{\bibfnamefont{A.}~\bibnamefont{Polls}}
  (\bibinfo{publisher}{Springer}, \bibinfo{address}{Berlin},
  \bibinfo{year}{1998}), vol. \bibinfo{volume}{510} of
  \emph{\bibinfo{series}{Lecture Notes in Physics}}, p. \bibinfo{pages}{269}.

\bibitem[{\citenamefont{Mitas}(1999)}]{mitas99}
\bibinfo{author}{\bibfnamefont{L.}~\bibnamefont{Mitas}}, in
  \emph{\bibinfo{booktitle}{Quantum {M}onte {C}arlo methods in physics and
  chemistry}}, edited by \bibinfo{editor}{\bibfnamefont{M.~P.}
  \bibnamefont{Nightingale}} \bibnamefont{and}
  \bibinfo{editor}{\bibfnamefont{C.~J.} \bibnamefont{Umrigar}}
  (\bibinfo{publisher}{NATO Advanced Study Institute on QMC},
  \bibinfo{address}{Cornell}, \bibinfo{year}{1999}), p. \bibinfo{pages}{247}.

\bibitem[{\citenamefont{Carlson}(1999)}]{carlson99b}
\bibinfo{author}{\bibfnamefont{J.}~\bibnamefont{Carlson}}, in
  \emph{\bibinfo{booktitle}{Quantum {M}onte {C}arlo methods in physics and
  chemistry}}, edited by \bibinfo{editor}{\bibfnamefont{M.~P.}
  \bibnamefont{Nightingale}} \bibnamefont{and}
  \bibinfo{editor}{\bibfnamefont{C.~J.} \bibnamefont{Umrigar}}
  (\bibinfo{publisher}{NATO Advanced Study Institute on QMC},
  \bibinfo{address}{Cornell}, \bibinfo{year}{1999}), p. \bibinfo{pages}{287}.

\bibitem[{\citenamefont{Pederiva et~al.}(2004)\citenamefont{Pederiva, Sarsa,
  Schmidt, and Fantoni}}]{pederiva04}
\bibinfo{author}{\bibfnamefont{F.}~\bibnamefont{Pederiva}},
  \bibinfo{author}{\bibfnamefont{A.}~\bibnamefont{Sarsa}},
  \bibinfo{author}{\bibfnamefont{K.~E.} \bibnamefont{Schmidt}},
  \bibnamefont{and} \bibinfo{author}{\bibfnamefont{S.}~\bibnamefont{Fantoni}},
  \bibinfo{journal}{Nucl. Phys. A} \textbf{\bibinfo{volume}{742}},
  \bibinfo{pages}{255} (\bibinfo{year}{2004}).

\bibitem[{\citenamefont{Carlson
  et~al.}(2003{\natexlab{b}})\citenamefont{Carlson, Chang, Pandharipande, and
  Schmidt}}]{carlson03c}
\bibinfo{author}{\bibfnamefont{J.}~\bibnamefont{Carlson}},
  \bibinfo{author}{\bibfnamefont{S.-Y.} \bibnamefont{Chang}},
  \bibinfo{author}{\bibfnamefont{V.~R.} \bibnamefont{Pandharipande}},
  \bibnamefont{and} \bibinfo{author}{\bibfnamefont{K.~E.}
  \bibnamefont{Schmidt}}, \bibinfo{journal}{Phys. Rev. Lett.}
  \textbf{\bibinfo{volume}{91}}, \bibinfo{pages}{050401}
  (\bibinfo{year}{2003}{\natexlab{b}}).

\bibitem[{\citenamefont{Chang et~al.}(2004)\citenamefont{Chang, Pandharipande,
  Carlson, and Schmidt}}]{chang04b}
\bibinfo{author}{\bibfnamefont{S.~Y.} \bibnamefont{Chang}},
  \bibinfo{author}{\bibfnamefont{V.~R.} \bibnamefont{Pandharipande}},
  \bibinfo{author}{\bibfnamefont{J.}~\bibnamefont{Carlson}}, \bibnamefont{and}
  \bibinfo{author}{\bibfnamefont{K.~E.} \bibnamefont{Schmidt}},
  \bibinfo{journal}{Phys. Rev. A} \textbf{\bibinfo{volume}{70}},
  \bibinfo{pages}{043602} (\bibinfo{year}{2004}).

\bibitem[{\citenamefont{Bouchaud et~al.}(1988)\citenamefont{Bouchaud, Georges,
  and Lhuillier}}]{bouchaud1988}
\bibinfo{author}{\bibfnamefont{J.~P.} \bibnamefont{Bouchaud}},
  \bibinfo{author}{\bibfnamefont{A.}~\bibnamefont{Georges}}, \bibnamefont{and}
  \bibinfo{author}{\bibfnamefont{C.}~\bibnamefont{Lhuillier}},
  \bibinfo{journal}{J. Physique} \textbf{\bibinfo{volume}{49}},
  \bibinfo{pages}{553} (\bibinfo{year}{1988}).

\bibitem[{\citenamefont{Fabrocini et~al.}(2005)\citenamefont{Fabrocini,
  Fantoni, Illarionov, and Schmidt}}]{fabrocini05}
\bibinfo{author}{\bibfnamefont{A.}~\bibnamefont{Fabrocini}},
  \bibinfo{author}{\bibfnamefont{S.}~\bibnamefont{Fantoni}},
  \bibinfo{author}{\bibfnamefont{A.~Y.} \bibnamefont{Illarionov}},
  \bibnamefont{and} \bibinfo{author}{\bibfnamefont{K.~E.}
  \bibnamefont{Schmidt}}, \bibinfo{journal}{Phys. Rev. Lett.}
  \textbf{\bibinfo{volume}{95}}, \bibinfo{pages}{192501}
  (\bibinfo{year}{2005}).

\bibitem[{\citenamefont{Bajdich et~al.}(2008)\citenamefont{Bajdich, Mitas,
  Wagner, and Schmidt}}]{bajdich08}
\bibinfo{author}{\bibfnamefont{M.}~\bibnamefont{Bajdich}},
  \bibinfo{author}{\bibfnamefont{L.}~\bibnamefont{Mitas}},
  \bibinfo{author}{\bibfnamefont{L.~K.} \bibnamefont{Wagner}},
  \bibnamefont{and} \bibinfo{author}{\bibfnamefont{K.~E.}
  \bibnamefont{Schmidt}}, \bibinfo{journal}{Phys. Rev. B}
  \textbf{\bibinfo{volume}{77}}, \bibinfo{pages}{115112}
  (\bibinfo{year}{2008}).

\bibitem[{\citenamefont{Fabrocini et~al.}(2008)\citenamefont{Fabrocini,
  Fantoni, Illarionov, and Schmidt}}]{fabrocini08}
\bibinfo{author}{\bibfnamefont{A.}~\bibnamefont{Fabrocini}},
  \bibinfo{author}{\bibfnamefont{S.}~\bibnamefont{Fantoni}},
  \bibinfo{author}{\bibfnamefont{A.~Y.} \bibnamefont{Illarionov}},
  \bibnamefont{and} \bibinfo{author}{\bibfnamefont{K.~E.}
  \bibnamefont{Schmidt}}, \bibinfo{journal}{Nucl. Phys. A}
  \textbf{\bibinfo{volume}{803}}, \bibinfo{pages}{137} (\bibinfo{year}{2008}).

\bibitem[{\citenamefont{Gezerlis and
  Carlson}(2008{\natexlab{b}})}]{gezerlis08b}
\bibinfo{author}{\bibfnamefont{A.}~\bibnamefont{Gezerlis}} \bibnamefont{and}
  \bibinfo{author}{\bibfnamefont{J.}~\bibnamefont{Carlson}},
  \bibinfo{journal}{private communication}
  (\bibinfo{year}{2008}{\natexlab{b}}).

\bibitem[{\citenamefont{Lagaris and Pandharipande}(1981)}]{lagaris81}
\bibinfo{author}{\bibfnamefont{I.~E.} \bibnamefont{Lagaris}} \bibnamefont{and}
  \bibinfo{author}{\bibfnamefont{V.~R.} \bibnamefont{Pandharipande}},
  \bibinfo{journal}{Nucl. Phys. A} \textbf{\bibinfo{volume}{359}},
  \bibinfo{pages}{331} (\bibinfo{year}{1981}).

\bibitem[{\citenamefont{Wambach et~al.}(1993)\citenamefont{Wambach, Ainsworth,
  and Pines}}]{wambach93}
\bibinfo{author}{\bibfnamefont{J.}~\bibnamefont{Wambach}},
  \bibinfo{author}{\bibfnamefont{T.~L.} \bibnamefont{Ainsworth}},
  \bibnamefont{and} \bibinfo{author}{\bibfnamefont{D.}~\bibnamefont{Pines}},
  \bibinfo{journal}{Nucl. Phys. A} \textbf{\bibinfo{volume}{555}},
  \bibinfo{pages}{128} (\bibinfo{year}{1993}).

\bibitem[{\citenamefont{Chen et~al.}(1993)\citenamefont{Chen, Clark, Dav\'e,
  and Khodel}}]{chen93}
\bibinfo{author}{\bibfnamefont{J.~M.~C.} \bibnamefont{Chen}},
  \bibinfo{author}{\bibfnamefont{J.~W.} \bibnamefont{Clark}},
  \bibinfo{author}{\bibfnamefont{R.~D.} \bibnamefont{Dav\'e}},
  \bibnamefont{and} \bibinfo{author}{\bibfnamefont{V.~V.}
  \bibnamefont{Khodel}}, \bibinfo{journal}{Nucl. Phys. A}
  \textbf{\bibinfo{volume}{555}}, \bibinfo{pages}{59} (\bibinfo{year}{1993}).

\bibitem[{\citenamefont{Schulze et~al.}(1996)\citenamefont{Schulze, Cugnon,
  Lejeune, Baldo, and Lombardo}}]{schulze96}
\bibinfo{author}{\bibfnamefont{H.~J.} \bibnamefont{Schulze}},
  \bibinfo{author}{\bibfnamefont{J.}~\bibnamefont{Cugnon}},
  \bibinfo{author}{\bibfnamefont{A.}~\bibnamefont{Lejeune}},
  \bibinfo{author}{\bibfnamefont{M.}~\bibnamefont{Baldo}}, \bibnamefont{and}
  \bibinfo{author}{\bibfnamefont{U.}~\bibnamefont{Lombardo}},
  \bibinfo{journal}{Phys. Lett. B} \textbf{\bibinfo{volume}{375}},
  \bibinfo{pages}{1} (\bibinfo{year}{1996}).

\bibitem[{\citenamefont{Schwenk et~al.}(2003)\citenamefont{Schwenk, Friman, and
  Brown}}]{schwenk03}
\bibinfo{author}{\bibfnamefont{A.}~\bibnamefont{Schwenk}},
  \bibinfo{author}{\bibfnamefont{B.}~\bibnamefont{Friman}}, \bibnamefont{and}
  \bibinfo{author}{\bibfnamefont{G.~E.} \bibnamefont{Brown}},
  \bibinfo{journal}{Nucl. Phys. A} \textbf{\bibinfo{volume}{713}},
  \bibinfo{pages}{191} (\bibinfo{year}{2003}).

\bibitem[{\citenamefont{Cao et~al.}(2006)\citenamefont{Cao, Lombardo, and
  Schuck}}]{cao06}
\bibinfo{author}{\bibfnamefont{L.~G.} \bibnamefont{Cao}},
  \bibinfo{author}{\bibfnamefont{U.}~\bibnamefont{Lombardo}}, \bibnamefont{and}
  \bibinfo{author}{\bibfnamefont{P.}~\bibnamefont{Schuck}},
  \bibinfo{journal}{Phys. Rev. C} \textbf{\bibinfo{volume}{74}},
  \bibinfo{pages}{064301} (\bibinfo{year}{2006}).

\bibitem[{\citenamefont{Margueron et~al.}(2008)\citenamefont{Margueron, Sagawa,
  and Hagino}}]{margueron08}
\bibinfo{author}{\bibfnamefont{J.}~\bibnamefont{Margueron}},
  \bibinfo{author}{\bibfnamefont{H.}~\bibnamefont{Sagawa}}, \bibnamefont{and}
  \bibinfo{author}{\bibfnamefont{K.}~\bibnamefont{Hagino}},
  \bibinfo{journal}{Phys. Rev. C} \textbf{\bibinfo{volume}{77}},
  \bibinfo{pages}{054309} (\bibinfo{year}{2008}).

\bibitem[{\citenamefont{Hebeler et~al.}(2007)\citenamefont{Hebeler, Schwenk,
  and Friman}}]{hebeler07}
\bibinfo{author}{\bibfnamefont{K.}~\bibnamefont{Hebeler}},
  \bibinfo{author}{\bibfnamefont{A.}~\bibnamefont{Schwenk}}, \bibnamefont{and}
  \bibinfo{author}{\bibfnamefont{B.}~\bibnamefont{Friman}},
  \bibinfo{journal}{Phys. Lett. B} \textbf{\bibinfo{volume}{648}},
  \bibinfo{pages}{176} (\bibinfo{year}{2007}).

\bibitem[{\citenamefont{Pandharipande and Wiringa}(1979)}]{pandharipande79}
\bibinfo{author}{\bibfnamefont{V.~R.} \bibnamefont{Pandharipande}}
  \bibnamefont{and} \bibinfo{author}{\bibfnamefont{R.~B.}
  \bibnamefont{Wiringa}}, \bibinfo{journal}{Rev. Mod. Phys.}
  \textbf{\bibinfo{volume}{51}}, \bibinfo{pages}{821} (\bibinfo{year}{1979}).

\bibitem[{\citenamefont{Wick}(1950)}]{wick50}
\bibinfo{author}{\bibfnamefont{G.~C.} \bibnamefont{Wick}},
  \bibinfo{journal}{Phys. Rev.} \textbf{\bibinfo{volume}{80}},
  \bibinfo{pages}{268} (\bibinfo{year}{1950}).

\bibitem[{\citenamefont{Cayley}(1849)}]{cayley1849}
\bibinfo{author}{\bibfnamefont{A.}~\bibnamefont{Cayley}},
  \bibinfo{journal}{Journal f{\"u}r die reine angewandte Mathematik}
  \textbf{\bibinfo{volume}{38}}, \bibinfo{pages}{93} (\bibinfo{year}{1849}),
  \bibinfo{note}{reprinted in {\em The collected mathematical papers of Arthur
  Cayley}, (Cambridge University Press, Cambridge, 1889), vol. 2, p. 19.}

\end{thebibliography}

\end{document}